\documentclass[aps,prc,twocolumn,superscriptaddress]{revtex4-1}

\usepackage{graphicx}
\usepackage{amsmath}
\usepackage{subfigure,dcolumn}
\usepackage{appendix}

\usepackage{CJKutf8}

\usepackage[colorlinks,
linkcolor=blue,
anchorcolor=blue,
urlcolor=blue,
citecolor=blue]{hyperref}
\begin{document}

\title{Nucleon-$\Delta$ elastic cross section in isospin-asymmetric nuclear medium with inclusion of scalar-isovector $\delta$ meson field 
}

\author{Manzi Nan \begin{CJK}{UTF8}{gbsn}\end{CJK}}
\affiliation{Institute of Modern Physics, Chinese Academy of Sciences, Lanzhou 730000, China}
\affiliation{School of Nuclear Science and Technology, University of Chinese Academy of Sciences, Beijing 100049, China}
\affiliation{School of Science, Huzhou University, Huzhou 313000, China}
\author{Pengcheng Li  \begin{CJK}{UTF8}{gbsn}\end{CJK}}
\email[Corresponding author, ]{lipch@zjhu.edu.cn}
\affiliation{School of Science, Huzhou University, Huzhou 313000, China}
\author{Wei Zuo  \begin{CJK}{UTF8}{gbsn}\end{CJK}}
\affiliation{Institute of Modern Physics, Chinese Academy of Sciences, Lanzhou 730000, China}
\affiliation{School of Nuclear Science and Technology, University of Chinese Academy of Sciences, Beijing 100049, China}
\author{Qingfeng Li  \begin{CJK}{UTF8}{gbsn}\end{CJK}}
\email[Corresponding author, ]{liqf@zjhu.edu.cn}
\affiliation{School of Science, Huzhou University, Huzhou 313000, China}
\affiliation{Institute of Modern Physics, Chinese Academy of Sciences, Lanzhou 730000, China}
	
\date{\today}
\begin{abstract}
\vspace{0.5cm} 
\noindent
\textbf{Abstract:} The production, dynamic evolution, and decay of $\Delta$ particles play a crucial role in understanding the properties of high baryon density nuclear matter in intermediate-energy heavy-ion collisions. 
In this study, the energy-, density-, and isospin-dependent nucleon-$\Delta$ elastic cross sections ($\sigma^{*}_{N \Delta}$) were studied within the framework of the relativistic Boltzmann-Uehling-Uhlenbeck transport theory,
in which the $\delta$ meson field is considered in addition to the $\sigma$, $\omega$, and $\rho$ meson fields. 
The results show that the $\delta$ and $\rho$ meson related exchange terms have a nonnegligible contribution to $\sigma^{*}_{N \Delta}$ compared to only considering the $\rho$ meson exchange terms, although there is a significant cancellation on the cross section among these meson exchange terms.  
In addition, owing to the different effects of the medium correction on the effective masses of neutrons, protons, and differently charged $\Delta$s, the individual $\sigma^{*}_{N \Delta}$ exhibits an ordered isospin-asymmetry ($\alpha$) dependence, and $\sigma^{*}_{n\Delta}$ and $\sigma^{*}_{p\Delta}$ have opposite $\alpha$ dependencies. 
Moreover, the $\alpha$ dependence of the ratio $R(\alpha)=\sigma^{*}(\alpha)/\sigma^{*}(\alpha=0)$ for $n\Delta$ reaction channels satisfies $n\Delta^{++}>n\Delta^{+}>n\Delta^{0}>n\Delta^{-}$, while for $p\Delta$, it satisfies $p\Delta^{-}>p\Delta^{0}>p\Delta^{+}>p\Delta^{++}$.
In addition, the results indicate that the isospin effect on $\sigma^{*}_{N \Delta}$, which is mostly caused by the isovector $\rho$ and $\delta$ meson fields, is still significant at densities up to three times the normal nuclear density.
Finally, a parametrization of the energy-, density-, and isospin-dependent $N\Delta$ elastic cross section is proposed based on the microscopic calculated results. Thus, the in-medium $\sigma^{*}_{N \Delta}$ in the energy range of $\sqrt{s}$=2.3$\sim$3.0 GeV can be properly described.

\vspace{0.5cm} 
\noindent
\textbf{Keywords:} isospin-asymmetric nuclear matter, nucleon-$\Delta$ elastic cross section, RBUU transport theory
\newline
\textbf{DOI:} 10.1088/1674-1137/add8fd

\end{abstract}

	
\maketitle
	

\section{Introduction}
The investigation of the properties of isospin-asymmetric nuclear matter under extreme conditions is a timely issue in both nuclear physics and nuclear astrophysics \cite{Danielewicz:2002pu,Li:2008gp,Sorensen:2023zkk,Li:2021thg}. 
It plays a crucial role in understanding the complex dynamic processes of heavy-ion collisions (HICs), nuclear structure, and formation and evolution of
dense stars, such as neutron stars\cite{Stoecker:1986ci,Lattimer:2000kb,Vretenar:2005zz}.
Over the past two decades, significant progress has been made in constraining the isospin-symmetric nuclear equation of state (EoS) at subnormal and  normal densities through theoretical calculations and comparisons with experimental nuclear data. However, its density-dependent behavior,   particularly in high-density regions, remains largely unclear, with uncertainty increasing  rapidly as density  rises\cite{Baldo:2016jhp,Russotto:2016ucm,Huth:2021bsp}. 
Furthermore, the construction of advanced radioactive beam facilities and new HICs experiments in them, including the High Intensity heavy ion Accelerator Facility (HIAF) in China \cite{Zhou:2022pxl}, the Facility for Antiproton and
Ion Research (FAIR) in Germany \cite{CBM:2016kpk}, the Beam Energy Scan (BES) and fixed target (FXT) programs at the Relativistic Heavy Ion Collider (RHIC) in the United States \cite{Chen:2024zwk},  and  the  Nuclotron-based  Ion  Collider  fAcility
(NICA) in Russia \cite{MPD:2022qhn}, is expected to open up new opportunities for experimental and theoretical investigations in the higher energy  and higher density EoS of isospin-asymmetric nuclear matter.

Charged-pion related observables are commonly used
as sensitive probes for investigating the high-density asymmetric nuclear EoS in HICs at intermediate energies and have attracted considerable attention in recent years \cite{Li:2005gfa,Xiao:2008vm,Xu:2013aza,Song:2015hua,Godbey:2021tbt,Luong:2024eaq,Li:2022icu,Li:2022iil}.  
However, predictions from different hadronic transport models for charged-pion related observables differ, especially at high densities. For instance, pion yields and ratios, as well as rapidity and transverse momentum distributions predicted by these models,  often lack consistency and fail to accurately reproduce experimental data \cite{SpiRIT:2020sfn,HADES:2020ver,TMEP:2023ifw}. 
In HICs, it is known that at intermediate energies,  the pions are mostly produced from the decay  of $\Delta$(1232) particles. Therefore, the production, evolution, and decay of $\Delta$ particles in the isospin-asymmetric nuclear medium are critical for accurately understanding and constraining the asymmetric nuclear EoS through experimental measurements and dynamic simulations \cite{Li:2016xix,Li:2017pis,Tong:2020dku, Godbey:2021tbt,Larionov:2001va,Oset:1987re}.

Regarding the cross sections of particle production, evolution, and decay used in the simulation of HICs, one can  usually  derive  them  from the  Brueckner theory \cite{Bohnet:1989dzk,Han:2022quc}, Dirac-Brueckner theory \cite{Li:1993ef,TerHaar:1987ce}, variational approach \cite{Pandharipande:1992zz}, as well one-boson-exchange model \cite{Cui:2018gbe,Huber:1994ee,Machleidt:1987hj,Larionov:2003av}.
They can also be parametrized from the comparison  of
theoretical calculations with experimental data \cite{Li:2018wpv,Li:2022wvu,Wang:2020xgk}. 
With the help of self-consistent relativistic BUU (RBUU) transport theory, the isospin-dependent in-medium nucleon-nucleon ($NN$) elastic cross section ($\sigma^{*}_{NN\rightarrow NN}$) has been systematically studied \cite{Li:2000sha,Li:2003vd}. 
For $\Delta$-related cross sections, such as $NN$ inelastic cross sections $\sigma^{*}_{NN \rightarrow N\Delta}$ (hard-$\Delta$ production), the soft-$\Delta$ production $\sigma^{*}_{N\pi \rightarrow \Delta}$, and $\Delta$ absorption $\sigma^{*}_{N\Delta \rightarrow NN}$ channels have been calculated within the framework of the RBUU approach in which the $\sigma$, $\omega$, $\rho$ and $\delta$ meson fields are considered \cite{Li:2017pis,Li:2016xix,Godbey:2021tbt}. 
The calculated results not only confirm that these cross sections are energy-, density-, and isospin-dependent, but also indicate that the $\delta$ meson field causes a splitting effect on the effective masses of nucleons and $\Delta$ particles, leading to splitting in the cross section of individual channels.

Recently, charged-pion yields from Au+Au collisions at several GeV energies have been measured by the STAR and HADES Collaborations \cite{HADES:2020ver,Luong:2024eaq}. 
Although these beam energies are currently too high to accurately investigate nuclear symmetry energy using pion-related observables, they provide more precise experimental data that help improve the theoretical description of pion production in HICs, thereby enabling more accurate constraints on the EoS of high-density nuclear matter. 
However, a significant mismatch remains between the
charged-pion yields calculated by various transport models and the experimentally measured values reported by the HADES collaboration \cite{HADES:2020ver}. By  considering an isospin-dependent reduction factor on  the $\Delta$ production, the charged-pion yields can be described properly \cite{Godbey:2021tbt,Kummer:2023hvl}.
To achieve a more accurate understanding of the
dense nuclear EoS, it is essential to compare measured pion-related observables with transport model simulation results. In addition to the channels of single single-$\Delta$ production and absorption, other channels (e.g., the $N\Delta$ elastic channels) should be self-consistently treated in the same transport model.

In a previous work \cite{Mao:1996zz}, the $N\Delta$ elastic cross section $\sigma^{*}_{N\Delta\rightarrow N\Delta}$ was calculated within the RBUU approach, only the isoscalar $\sigma$ and $\omega$ meson exchanges were involved. 
Then, the isovector $\rho$ meson exchange was further considered to investigate the contribution of the isovector field on $\sigma^{*}_{N\Delta\rightarrow N\Delta}$\cite{Nan:2023gwp}. 
In the relativistic mean field theory, the bulk properties of nuclei, such as binding energy and charge radius, can be precisely predicted by introducing the isovector $\rho$ meson field \cite{Long:2006nc,Long:2010qc}. 
In addition, it has been pointed out that the $\delta$ meson field plays a crucial role in accurately describing strongly isospin asymmetric matter at high densities in neutron stars, 
directly affecting the density dependence of the symmetry energy and causing a splitting of the Dirac mass for protons and neutrons in asymmetric matter.
\cite{Kubis:1997ew,Li:2016xix,Dutra:2014qga,Santos:2024aii}.
For instance, Ref.~\cite{Roca-Maza:2011alv} demonstrated that the inclusion of the \(\delta\) meson field not only improves the accuracy of mass and radius predictions for finite nuclei but also influences the EoS at higher densities, resulting in much better agreement with heavy-ion collision data.  
Furthermore, the \(\sigma^2\delta^2\) mixing terms in the nonlinear coupling of the effective Lagrangian significantly affect astrophysical observables, such as the radius and tidal deformability of neutron stars \cite{Miyatsu:2022wuy}.

In this study, based on the effective Lagrangian within the same framework of the RBUU microscopic transport theory, in which the scalar-isovector $\delta$ meson exchange is considered, we further studied energy-, density-, and isospin-dependent $N\Delta \rightarrow N\Delta$ cross section more systematically.

This paper is organized as follows. 
A brief review of the RBUU equation and analytic expressions of in-medium $N\Delta \rightarrow N\Delta$ cross sections are presented in Sec.\ref{sec2}.  
Numerical results of total and individual $\sigma^{*}_{N\Delta}$, as well as the effective mass splitting effects on the cross section are presented in Sec.\ref{sec3}. 
Conclusion and outlook are provided in Sec.\ref{sec4}.

\section{Formulation}\label{sec2}
The same theoretical framework as that established in Refs. \cite{Li:2000sha,Li:2017pis,Li:2003vd,Mao:1996zz} was employed in this study.
By using the closed time-path Green's function technique, which is extensively employed to process issues related to non-equilibrium systems \cite{Buss:2011mx}, and incorporating the semi-classical and quasi-particle approximations, the RBUU equation for the $\Delta$ distribution function can be derived as \cite{Mao:1996zz}
\begin{equation}
\begin{aligned}
    &\left\{p_{\mu}\left[\partial_{x}^{\mu}-\partial_{x}^{\mu} \Sigma_{\Delta}^{v}(x) \partial_{v}^{p}+\partial_{x}^{v} \Sigma_{\Delta}^{\mu}(x) \partial_{v}^{p}\right]+m_{\Delta}^{*} \partial_{x}^{v} \Sigma_{\Delta}^{S}(x) \partial_{v}^{p}\right\} \\
    &\frac{f_{\Delta}(\mathbf{x}, \mathbf{p}, \tau)}{E_{\Delta}^{*}(p)}=C^{\Delta}(x, p).
    \label{eq.1}
\end{aligned}
\end{equation}
Here, $m_{\Delta}^{*}$ and $f_{\Delta}(\mathbf{x}, \mathbf{p}, \tau)$ represent the effective mass and distribution function of $\Delta(1232)$, respectively. $\Sigma_{\Delta}^{S}$ and $\Sigma_{\Delta}^{\mu,\nu}$ on the left side characterize the Hartree terms of the $\Delta$ self-energies. $C^{\Delta}(x, p)$ on the right side represents the collision term, which is determined by the collisional self-energy and is closely related to the in-medium elastic and inelastic cross sections.

In the present study, we exploratorily introduce the scalar-isovector $\delta$ meson field in the effective Lagrangian, along with the scalar-isoscalar $\sigma$, the vector-isoscalar $\omega$, and vector-isovector $\rho$ meson fields, with the aim of understanding the impact of including the $\delta$ meson field on the description of $N \Delta \rightarrow N \Delta$ scattering. 
It should be noted that $N\Delta$ elastic cross sections in free space can be understood primarily with the help of $\pi$ meson exchange. This is due to the long-range nuclear exchange characteristics of $\pi
$ meson, which effectively provide cross section within free space showing good agreement with Cugnon’s parametrization in the higher energy region. However, in the nuclear medium, other meson exchanges, such as $\sigma$, $\omega$, $\rho$, and $\delta$, become increasingly significant and dominant at higher densities. 

Thus, the effective Lagrangian can be written as
\begin{equation}
L=L_{F}+L_{I},
\end{equation}
where $L_{F}$ is the free Lagrangian density and $L_{I}$ is for the interaction part,
\begin{equation}
    \begin{aligned}
L_{F}= & \bar{\Psi}\left[i \gamma_{\mu} \partial^{\mu}-m_{N}\right] \Psi+\bar{\Psi}_{\Delta \nu}\left[i \gamma_{\mu} \partial^{\mu}-m_{\Delta}\right] \Psi_{\Delta}^{\nu} \\
& +\frac{1}{2} \partial_{\mu} \sigma \partial^{\mu} \sigma+\frac{1}{2} \partial_{\mu} \vec{\delta} \partial^{\mu} \vec{\delta}-\frac{1}{4} F_{\mu \nu} \cdot F^{\mu v}-\frac{1}{4} \vec{L}_{\mu\nu} \cdot \vec{L}^{\mu \nu} \\
& -\frac{1}{2} m_{\sigma}^{2} \sigma^{2}-\frac{1}{2} m_{\delta}^{2} \vec{\delta}^{2} +\frac{1}{2} m_{\omega}^{2} \omega_{\mu} \omega^{\mu}+\frac{1}{2} m_{\rho}^{2} \vec{\rho}_{\mu} \vec{\rho}^{\mu},
\end{aligned}
\end{equation}
\begin{equation}
    \begin{aligned}
L_{I}= & g_{N N}^{\sigma} \bar{\Psi} \Psi \sigma+g_{N N}^{\delta} \bar{\Psi}\vec{\tau }\cdot \Psi \vec{\delta} -g_{N N}^{\omega} \bar{\Psi} \gamma_{\mu} \Psi \omega^{\mu}\\
& -g_{N N}^{\rho} \bar{\Psi} \gamma_{\mu} \vec{\tau } \cdot \Psi \vec{\rho }^{\mu} +g_{\Delta \Delta}^{\sigma} \bar{\Psi}_{\Delta} \Psi_{\Delta} \sigma+g_{\Delta \Delta}^{\delta} \bar{\Psi}_{\Delta} \vec{\tau } \cdot \Psi_{\Delta} \vec{\delta} \\
&-g_{\Delta \Delta}^{\omega} \bar{\Psi}_{\Delta} \gamma_{\mu} \Psi_{\Delta} \omega^{\mu}-g_{\Delta \Delta}^{\rho} \bar{\Psi}_{\Delta} \gamma_{\mu} \vec{\tau} \cdot \Psi_{\Delta} \vec{\rho }^{\mu},
\end{aligned}
\end{equation}
where $F_{\mu \nu} \equiv \partial_{\mu} \omega_{v}-\partial_{v} \omega_{\mu} , L_{\mu \nu} \equiv \partial_{\mu} \vec{\rho}_{v}-\partial_{v} \vec{\rho}_{\mu}$, $\psi$ is the Dirac spinor, and $\psi_{\triangle}$ is the Rarita-Schwinger spinor.

In this study, we adopted density-dependent coupling constants, which have been extensively applied in the calculation of both elastic and inelastic reaction channels. As a result, we provide a more accurate description of cross sections for $NN\rightarrow NN$, $NN\rightarrow N\Delta$, and $N\pi \rightarrow\Delta$\cite{Li:2016xix,Li:2003vd,Li:2017pis,Hofmann:2000vz}. Thus, it can be quantitatively parametrized as
\begin{equation}
\label{eq.5}
		g_{q}(\rho_{b})=g_{q}(\rho_{0})f_{q}(u), \quad  q=\sigma,~\omega,~\rho,~\delta
	\end{equation}
 where $u=\rho_{b}/\rho_{0}$, $\rho_{b}$ and $\rho_{0}$ are the baryon and normal nuclear densities, respectively, $f_{q}(u)$ reads as
	\begin{equation}
		f_{q}(u)=a_{q}\dfrac{1+b_{q}(u+d_{q})^{2}}{1+c_{q}(u+d_{q})^{2}}.
	\end{equation}

For the $\Delta$-$\Delta$-meson vertex, the coupling constant ratio is defined as $\chi_{q} = g^{q}_{\Delta \Delta}/g^{q}_{NN}$. We set $\chi_{\sigma}=1.0$, $\chi_{\omega}=0.8$ in this study, these parameter values lie within the parameter space, which is obtained by comparing theoretical investigations and experimental data \cite{Li:2018qaw,Kosov:1998gp,Sun:2018tmw,Drago:2014oja,Wehrberger:1989cd}.  
Regarding $\chi_\rho$ and $\chi_\delta$, they have not been strictly constrained through comparison of theoretical predictions and experimental data. 
In Ref.\cite{Raduta:2021xiz}, $0.7 \leq \chi_\rho \leq 1.3$ was adopted to investigate the $\Delta-$admixed neutron stars, and it was shown that for a large domain of the parameter space, nucleation of $\Delta$s opens-up the nucleonic dUrca process which is otherwise forbidden.
In this study, we adopted a fixed value of $\chi_\delta$ = $\chi_\rho=0.7$ for simplicity. It is noted that the variation in $\chi_\rho$ and $\chi_\delta$ within 0.7$\sim$1.3 introduces some uncertainty in the calculated cross sections. However, the main conclusions of this investigation are not affected by the choice of this parameter.

According to the relativistic mean field theory, the effective masses of the nucleons and $\Delta$ particles are related to the average value of the meson fields and have the following forms\cite{Song:2015hua}: 
\begin{equation}
    \begin{array}{l}
m_{p / n}^{*}=m_{N}-g_{\sigma} \sigma \mp g_{\delta} \delta_{0}, \\
m_{\Delta^{++}/ \Delta^{-}}^{*}= m_{\Delta} -g_{\sigma} \sigma \mp g_{\delta} \delta_{0}, \\
m_{\Delta^{+} / \Delta^{0}}^{*}= m_{\Delta} -g_{\sigma} \sigma \mp \frac{1}{3} g_{\delta} \delta_{0}.
\end{array}
\label{eq.7}
\end{equation}
Here, the nucleon mass \(m_{N}\) in free space is taken as 0.938 GeV,
and $m^{*}_{p}$, $m^{*}_{n}$, $m^{*}_{\Delta^{++}}$, $m^{*}_{\Delta^{+}}$, $m^{*}_{\Delta^{0}}$, and $m^{*}_{\Delta^{-}}$ represent the effective masses of the proton, neutron, $\Delta^{++}$, $\Delta^{+}$, $\Delta^{0}$, and $\Delta^{-}$ in the nuclear medium, respectively.  
The coupling constants $g_{\sigma}$, $g_{\omega}$, $g_{\rho}$, $g_{\delta}$ 
are derived from Eq. \ref{eq.5}. The values of the $\sigma$ and $\delta$ meson fields are determined by solving the corresponding Klein-Gordon equation. 
In neutron-rich matter, the isospin asymmetry parameter is defined as $\alpha=(\rho_{n}-\rho_{p})/(\rho_{n}+\rho_{p})\neq0$, and the effective masses of nucleons and $\Delta$ particles obey \(m^{*}_{p} > m^{*}_{n}\) and \(m^{*}_{\Delta^{++}} > m^{*}_{\Delta^{+}} > m^{*}_{\Delta^{0}} > m^{*}_{\Delta^{-}}\) \cite{Li:2017pis}.

The collision term in Eq.~\ref{eq.1} can be divided into the $\Delta$-related elastic, inelastic, and decay interaction parts 
\cite{Li:1991pq,Wang:1991sj,Mao:1996zz},
\begin{equation}
     C^{\Delta}(x, p)=C^{\Delta}_{\text{el}}(x, p)+C^{\Delta}_{\text{in}}(x,p)+C^{\Delta}_{N\pi}(x, p),
\end{equation}
and the elastic part can be further distinguished into $N\Delta$ and $\Delta\Delta$ elastic interaction parts, 
\begin{equation}
    C^{\Delta}_{\text{el}}(x, p)=C^{N\Delta}_{\text{el}}(x, p)+C^{\Delta\Delta}_{\text{el}}(x, p).
\end{equation}
In this work, we focused exclusively on \( N\Delta \) elastic scattering, and it can be expressed as
\begin{equation}
\begin{split}
 C^{N\Delta}_{\text{el}}(x, p) &= \frac{1}{4} \int \frac{d^{} \mathbf{p}_{2}}{(2 \pi)^{3}} \int \frac{d^{} \mathbf{p}_{3}}{(2 \pi)^{3}} \int \frac{d^{} \mathbf{p}_{4}}{(2 \pi)^{3}}  \\
&  \times(2 \pi)^{4} \delta^{(4)}(p_{1}+p_{2} -p_{3}-p_{4})\\
& \times W^{N\Delta}_{\text{el}}(p_{1}, p_{2}, p_{3}, p_{4})[F_{2}-F_{1}] \\
&=\frac{1}{4} \int \frac{d \mathbf{p}_{2}}{(2 \pi)^{3}} \sigma^{N\Delta}_{\text{el}}(s, t) v_{\Delta}\left[F_{2}-F_{1}\right] d \Omega,
\label{eq.8}
\end{split}
\end{equation}
where $\sigma^{N\Delta}_{\text{el}}(s,t)$ denotes the $N \Delta \rightarrow N \Delta$ cross section, and $F_{2}$ and $F_{1}$ are the Uehling-Uhlenbeck Pauli-blocking factors of the loss and gain terms. 
The transition probability in $N\Delta\rightarrow N\Delta$ can be expressed as
\begin{equation}
    W^{N\Delta}_{\text{el}}\left(p, p_{2}, p_{3}, p_{4}\right)=G\left(p, p_{2}, p_{3}, p_{4}\right)+p_{3} \leftrightarrow p_{4},
\end{equation}
and
\begin{equation}
    G=\frac{g_{\Delta \Delta}^{A} g_{\Delta \Delta}^{B} g_{N N}^{A} g_{N N}^{B}}{16 E_{\Delta}^{*}(p) E^{*}\left(p_{2}\right) E_{\Delta}^{*}\left(p_{3}\right) E^{*}\left(p_{4}\right)}  T_{e} \Phi_{e},
    \label{eq.11}
\end{equation}
where $T_{e}$ and $\Phi_{e}$ are the isospin and the spin matrices, respectively;
the terms $g^{A, B}_{\Delta \Delta}, g^{A, B}_{N N}$ are the coupling constants for $\Delta$-$\Delta$-meson and nucleon-nucleon-meson interactions, respectively, and ($A, B$) denotes the type of meson exchanges involved.

The individual differential cross sections read as
\begin{equation}
\frac{d \sigma_{N \Delta \rightarrow N \Delta}^{*}}{d \Omega}=\frac{1}{(2 \pi)^{2} s} \sum_{r=1}^{10} \frac{1}{32}\left[d_{r} D_{r}(s, t)+(s, t \leftrightarrow u)\right],
\end{equation}
The indices $r=1$ to $10$ respectively correspond to the meson exchange terms: $\sigma-\sigma$, $\omega-\omega$, $\sigma-\omega$, $\rho-\rho$, $\delta-\delta$, $\delta-\rho$, $\sigma-\delta$, $\sigma-\rho$, $\omega-\delta$, and $\omega-\rho$. Detailed information about the parameters $d_{i}$ and $D_{i}$ can be found in Tab.~\ref{table 1-1} and App. \ref{App_A}.  
Regarding the $d_{i}$ component of the total cross section, it is necessary to average the isospin matrix for each individual channel, as shown in the last row of Tab.~\ref{table 1-1}. 

In addition, we used the following phenomenological effective form factor for the nucleon-nucleon-meson vertex, owing to the finite size and short-range correlation properties of baryons:
\begin{equation}
F_{q}(t)= \frac{\Lambda_{q}^{2}}{\Lambda_{q}^{2}-t}.
\label{eq.12}
\end{equation}
The cutoff masses for different mesons, denoted as $\Lambda_{q}$, were taken as $\Lambda_{\sigma}=1.1$ GeV, $\Lambda_{\omega}=0.783$ GeV, $\Lambda_{\rho}=0.770$ GeV, $\Lambda_{\delta}=0.983$ GeV \cite{Li:2003vd,Li:2000sha}, 
and $\Lambda_{\Delta}=0.4\Lambda$, consistent with the value chosen in Ref.~\cite{Nan:2023gwp}.

Moreover, other factors that might affect $N\Delta$ scattering should be mentioned here, such as the canonical momenta correction and threshold effect. In this study, we primarily focus on the isospin dependence of $N\Delta$ elastic cross section at densities $u \leq 3\rho_{0}$. Above this density, the quasi-particle approximation used in this transport theory may become unreliable, as it cannot adequately describe the strong modifications to baryon properties under extreme conditions. Additionally, a possible phase transition from hadronic gas to quark-gluon-plasma (QGP) may occur \cite{Almaalol:2022xwv,Li:2022cfd}. 
Although the above factors have some influence on the production and absorption of $\Delta$ and pions as well as on the charged-pion ratio \cite{Song:2015hua,Zhang:2017mps}, and should be considered carefully, their combined effects would complicate the conclusions of this study and will instead be consistently accounted for in numerical microscopic transport model simulations.

\begin{table}[!t]
	\centering
            \caption
            {Isospin matrix parameter sets $T_{e}$ for individual $N\Delta \rightarrow N\Delta$ reaction channels.}
		\label{table 1-1}
        \renewcommand{\arraystretch}{1.5} 
		\setlength{\tabcolsep}{2.3mm}{
		\begin{tabular}{l c c c}\hline\hline
		    & $\sigma-\sigma$, $\omega-\omega$  & $\delta-\delta$, $\rho-\rho$, &  $\omega-\rho$, $\sigma-\delta$,  \\
                 $              $ & $\sigma-\omega$  &  $\delta-\rho$  & $\sigma-\rho$, $\omega-\delta$ \\
		\hline
		$p\Delta^{++}(n\Delta^{-})$  & 1  & 9/4 & 3/2 \\
		
		$n\Delta^{++}(p\Delta^{-})$  & 1  & 9/4 & -3/2 \\
		
		$p\Delta^{+}(n\Delta^{0})$  & 1  & 1/4 &  1/2 \\
		
		$n\Delta^{+}(p\Delta^{0})$  & 1  & 1/4 & -1/2 \\
            $N\Delta\rightarrow N\Delta $  & 1  & 5/4 & 0 \\
		\hline \hline
\end{tabular}}
\end{table}

\begin{figure}[b]
    \centering
    \includegraphics[width=1\linewidth]{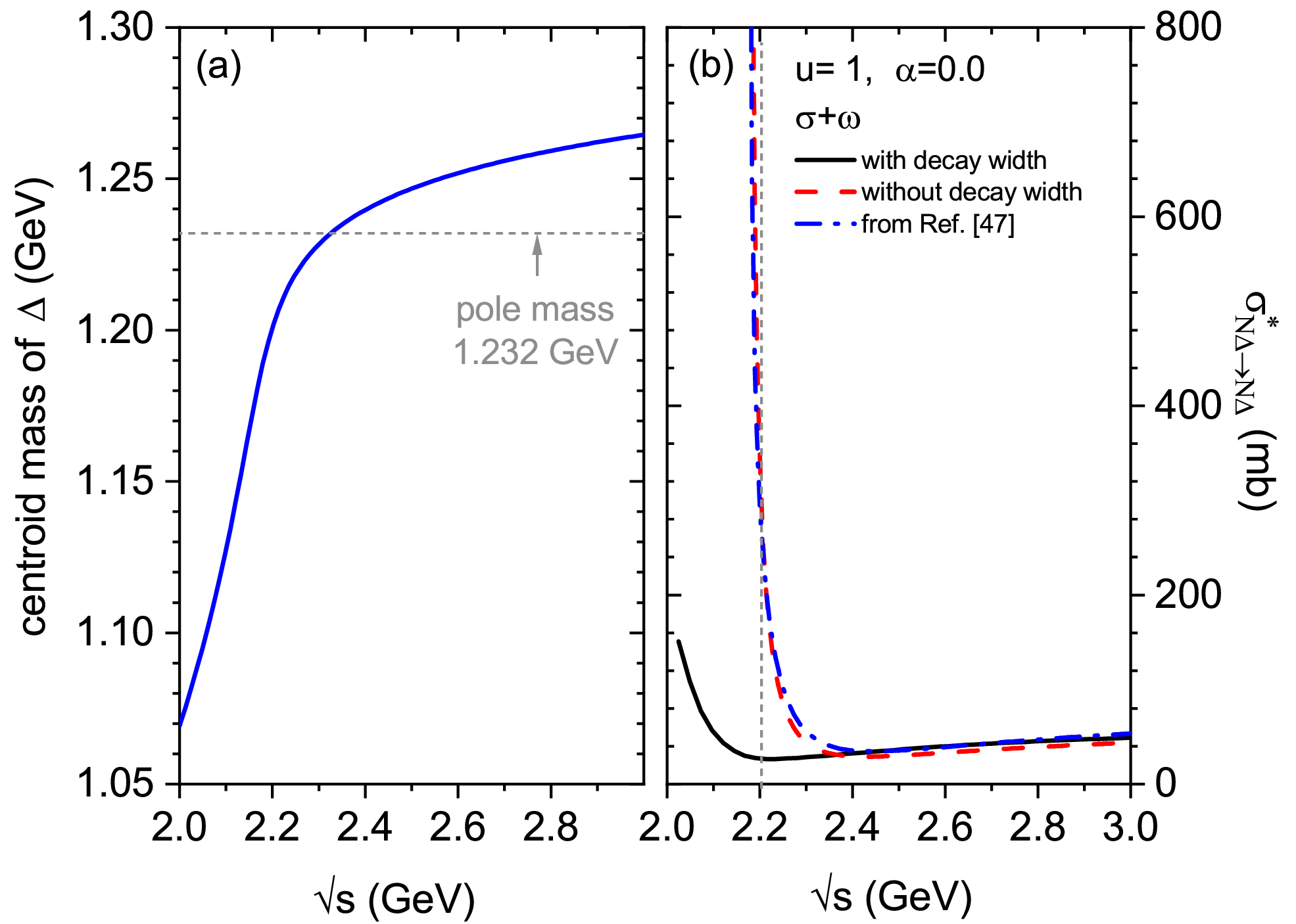}
   \caption{(Color online) Panel (a): the blue solid line represents the centroid mass of $\Delta(1232)$ as a function of center of mass (c.m.) energy, whereas the gray dashed line represents the pole mass of $\Delta$. 
   Panel (b): isospin-independent $N\Delta \rightarrow N\Delta$ cross sections with (black solid line) and without (red dash line) considering the $\Delta$ resonance decay width for $\alpha$=0 at $u$=1. The blue dashed-dot line represents the result from Ref. \cite{Nan:2023gwp} without considering the $\Delta$ resonance decay width.}
    \label{fig.1}
\end{figure}
\section{Results and discussion}\label{sec3}

\begin{figure}[b]
    \centering
    \includegraphics[width=1\linewidth]{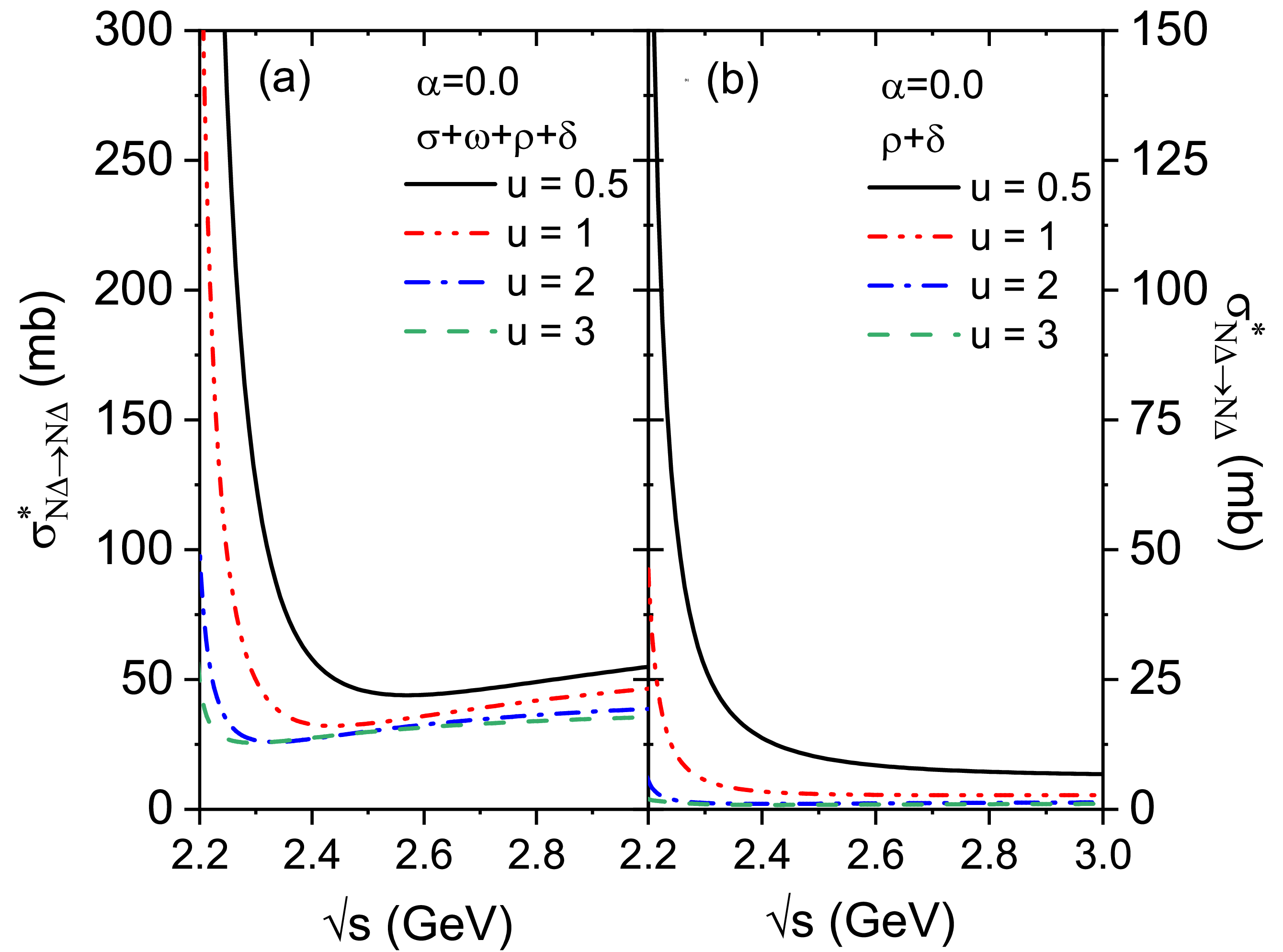}
    \caption{(Color online) $N\Delta\rightarrow N\Delta$ cross section as a function of c.m. energy for symmetric nuclear matter ($\alpha=0$) at $u$=0.5, 1, 2, and 3. 
    Panel (a) and (b) are calculated with the inclusion of the $\sigma+\omega+\rho+\delta$ and $\rho+\delta$ meson exchanges, respectively.}
   \label{fig.3a}
\end{figure}

First, Given that the $\Delta$ particle is an unstable resonance state in nature, it is important to investigate the influence of its decay width on the $N \Delta \rightarrow N \Delta$ cross section.  
Commonly, the decay width of $\Delta$-isobar can be calculated applying the quantum field theory or determined by the widely used momentum-dependent phenomenological formula \cite{Mao:1994zza,Zhang:2017mps,Li:2022icu}.
In the calculations of $NN\rightarrow N \Delta$ and $N\pi \rightarrow \Delta$ cross sections, their dependence on the decay width of $\Delta$ is accounted for by introducing the Breit-Wigner distribution function integral \cite{Li:1995pra,Li:2017pis,Song:2015hua}. 
To estimate the effect of the decay width of $\Delta$ on the $N \Delta \rightarrow N \Delta$ cross section, we used the centroid mass of $\Delta$ proposed in Refs.~\cite{Mao:1994zza,Mao:1996zz}. 
The energy dependence of the centroid mass of $\Delta$ is shown in Fig. \ref{fig.1}(a) represented by the blue solid line, while the gray dashed line represents the value of the $\Delta(1232)$ pole mass. The centroid mass of $\Delta$ increases rapidly with the center of mass (c.m.) energy at energies below approximately 2.2 GeV and then slows down noticeably.

Figure~\ref{fig.1}(b) shows $N\Delta \rightarrow N\Delta$ cross sections, which only include the contributions of $\sigma$ and $\omega$ meson related exchanges, with (black solid line) and without (red dashed line) considering the $\Delta$ resonance decay width for $\alpha$=0 at the reduced density $u$=1. 
The results shown in Fig.2(a) of Ref. \cite{Nan:2023gwp} are also represented by the blue dashed-dot line for comparison. 
Although the coupling constants used in the effective Lagrangian density differ, the calculation results in this study are similar because both sets of coupling constants were obtained by fitting the properties of finite nuclei.
Furthermore, when the $\Delta$ resonance decay width is considered, a significant suppression effect on $\sigma^{*}_{N\Delta}$ at lower energies ($\sqrt{s}\lesssim$ 2.2 GeV, left of the vertical gray dashed line) can be observed, while at higher energies ($\sqrt{s}\gtrsim$~2.2 GeV), the cross section becomes weakly dependent of the $\Delta$ resonance decay width.  
Therefore, in the following, we adopt the pole mass of the $\Delta$ particle simplicity and mainly focus on the $N\Delta \rightarrow N\Delta$ cross section at energies above 2.2 GeV. In addition, we systematically analyze the isospin and density dependences of the $N\Delta \rightarrow N\Delta$ cross sections at $\sqrt{s}=$2.5.

\begin{figure}[b]
    \centering
    \includegraphics[width=1\linewidth]{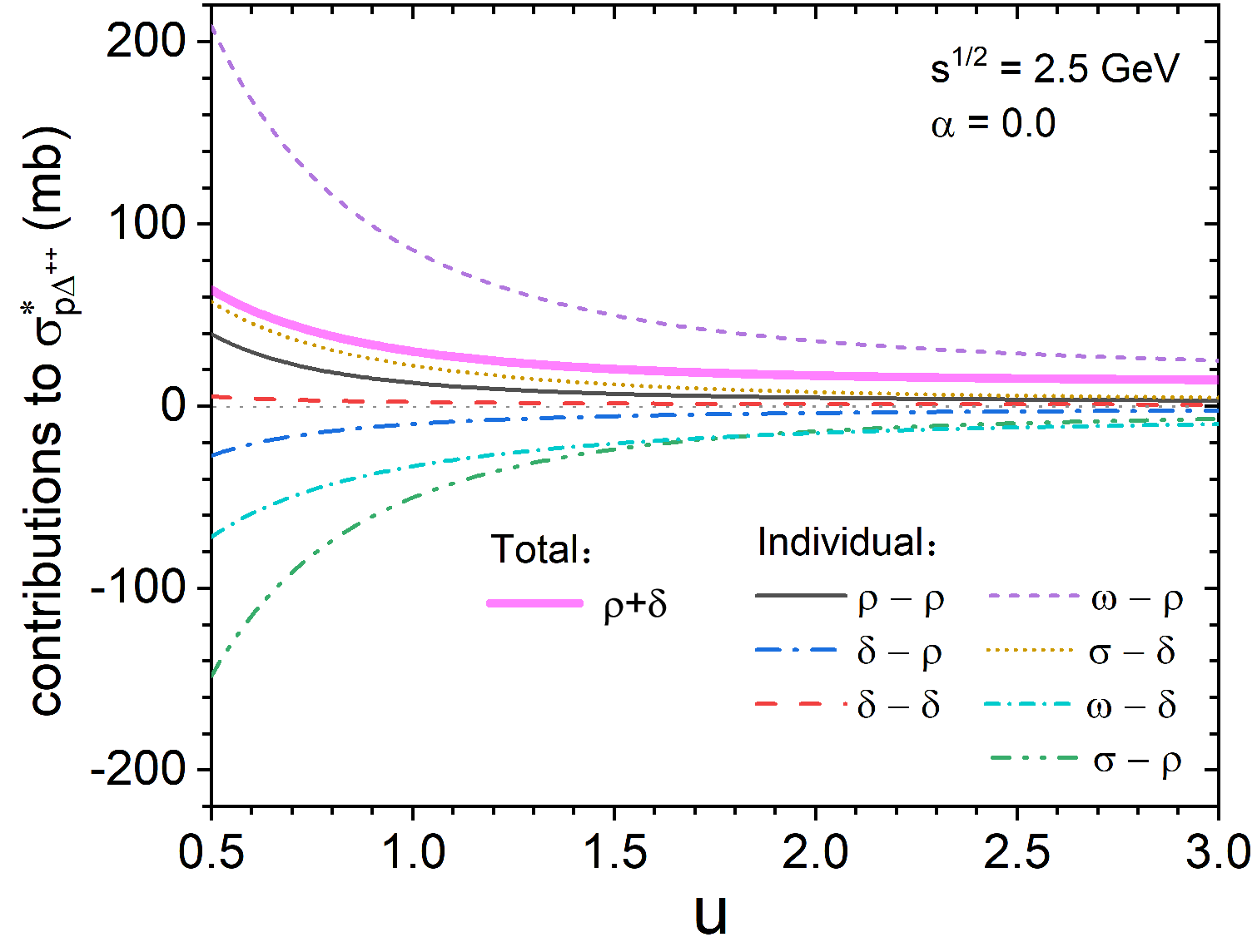}
    \caption{(Color online) Total (thick solid magenta line) and individual contributions of the $\rho$ and $\delta$ related exchange terms to $\sigma^{*}_{p\Delta^{++}}$ for $\alpha=0$ at $\sqrt{s}=$2.5 GeV. The horizontal gray dotted line represents zero.}
    \label{fig:cc33}
\end{figure}

\subsection{Density dependence of $\sigma^{*}_{N\Delta}(\sqrt{s},u)$}

\begin{figure*}[htbp!]
    \centering
    \includegraphics[width=1.0\linewidth]{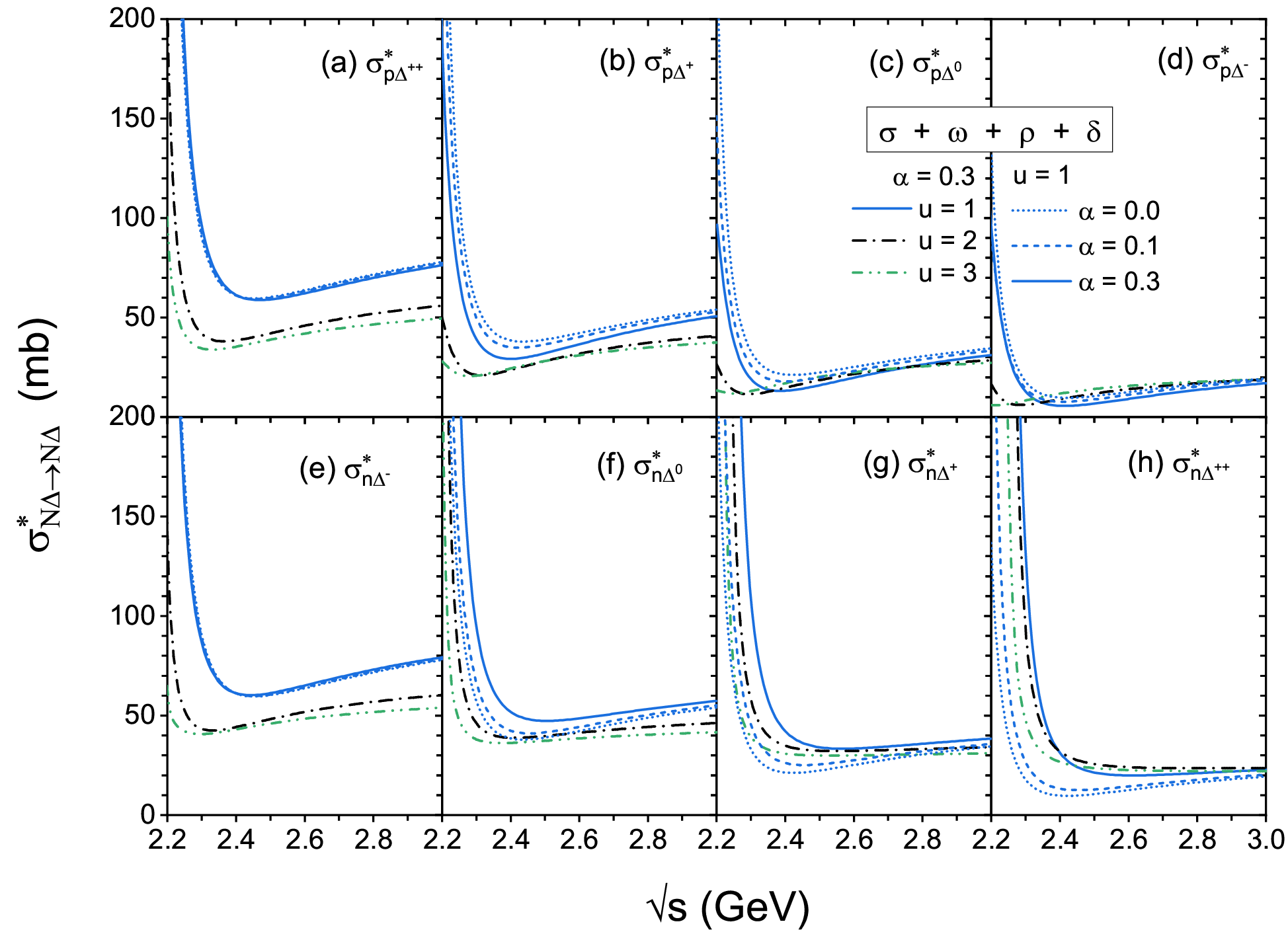}
    \caption{(Color online) Plots of $\sigma^{*}_{N\Delta\rightarrow N \Delta}$ as a function of c.m. energy at various densities ($u$=1,~2,~3) with isospin asymmetry degree ($\alpha$=0.0,~0.1,~0.3). The top and bottom panels correspond to individual $p\Delta$ and $n\Delta$ elastic cross sections, respectively.}
    \label{fig:graph3}
\end{figure*}

Fig. \ref{fig.3a}(a) shows the total contributions of $\sigma$, $\omega$, $\rho$, and $\delta$ meson exchanges to the energy- and density-dependent $\sigma^{*}_{N\Delta}(\sqrt{s},u)$ at various densities ($u$=0.5, 1, 2, and 3) for $\alpha=0$. 
Similar to the values of $\sigma^{*}_{NN\rightarrow N\Delta}$ reported in Refs. \cite{Song:2015hua,Li:2016xix}, and those of $\sigma^{*}_{N\Delta}$ (only include $\sigma$, $\omega$ and $\rho$ meson exchanges) reported in Ref. \cite{Nan:2023gwp}, the values of $\sigma^{*}_{N\Delta}$ shown in Fig. \ref{fig.3a}(a) decrease with increasing reduced density, indicating a visible density dependent suppression of nuclear medium on this cross section, especially at energies below approximately 2.4 GeV. This is mainly caused by the decrease in the  baryon effective mass with increasing density \cite{Li:2016xix}. 
When the energies exceed 2.4 GeV, the density dependence of the cross section weakens, and the cross section is slightly enhanced with increasing energy. According to the Walecka model, the scalar $\sigma$ and vector $\omega$ meson fields contribute to an attractive and a repulsive potential, respectively. 
Furthermore, the relative momentum between the ingoing nucleon and $\Delta$ increases with $\sqrt{s}$, and the momentum-dependent repulsion might be dominant and exert a greater effect, resulting in a possible increase in $\sigma^{*}_{N\Delta}$.

To clearly how the total effects of the isovector meson fields on $\sigma^{*}_{N\Delta}(\sqrt{s},u)$, Fig. \ref{fig.3a}(b) depicts the contributions of $\rho$ and $\delta$ meson involved terms (including the crossing terms with $\sigma$, $\omega$) to the $\sigma^{*}_{N\Delta \rightarrow N \Delta}$ at $u$=0.5, 1, 2, and 3 when $\alpha$=0.
When compared with Fig. \ref{fig.3a}(a), it can be observed that the proportion of the contributions of $\rho$ and $\delta$ meson involved terms in total $\sigma^{*}_{N \Delta \rightarrow N \Delta}$ is more pronounced at lower densities and suppressed as density increases. 
When further comparing with $\sigma^{*}_{N \Delta \rightarrow N \Delta}$, for which only the $\rho$ meson involved terms contribute, as shown in Fig.2(c) of Ref. \cite{Nan:2023gwp}, it can be concluded that the $\rho$ and $\delta$ meson related-terms have a larger contribution than that of the $\rho$ meson field alone.

It is interesting to further explore the individual contribution of each isovector meson related exchange to $\sigma^{*}_{N \Delta}(\sqrt{s},u)$. 
As shown in Tab. \ref{table 1-1}, the isospin matrix  of isospin vector-vector meson exchanges in Eq. \ref{eq.11} is 5/4, while that of isospin scalar-vector meson exchanges is 0 for the total $N \Delta \rightarrow N \Delta$ channel. Therefore, for the contributions of $\rho$ and $\delta$ related exchanges to the total $\sigma^{*}_{N \Delta}$, only the isospin vector-vector meson exchanges are taken into account. 
However, for individual channels, the isospin matrixes of isospin scalar-vector and vector-vector meson exchanges are different. Thus the total contribution of $\rho$ and $\delta$ related exchange terms to $\sigma^{*}_{p\Delta^{++}}$, which is represented by the thick solid magenta line in Fig. \ref{fig:cc33}, is larger than that to $\sigma^{*}_{N\Delta}$ shown in Fig. \ref{fig.3a}(b). 
In addition, the individual contributions of $\rho$ and $\delta$ related exchanges to each channel deserve to be further studied. 
Taking the $\sigma^{*}_{p \Delta^{++}}$ for $\alpha$=0 at $\sqrt{s}$=2.5 GeV as an example, the individual contributions of $\rho$ and $\delta$ meson related exchange are shown in Fig. \ref{fig:cc33}.
It can be found that the contribution of each meson exchange term to $\sigma^{*}_{p \Delta^{++}}$ decreases with increasing reduced density. This density dependence originates from the baryon-baryon-meson coupling constants and effective masses of nucleons and $\Delta$ particles. 
This density-dependent characteristic of the cross section suggests that, in addition to the linear term, an exponential term should be introduced in the parametrized formula, as shown in Eq. \ref{formul} and discussed in Sec. \ref{B3}. 
In addition, there exists an evident cancellation effect between $\omega-\rho$ and $\sigma-\rho$, $\sigma-\delta$ and $\omega-\delta$, $\rho-\rho$ and $\delta-\rho$, $\delta-\delta$ and $\delta-\rho$, respectively. However, the absolute values of the contributions of $\omega-\rho$, $\sigma-\delta$, $\rho-\rho$, and $\delta-\delta$ meson exchange terms to $\sigma^{*}_{p \Delta^{++}}$ are larger than those of the corresponding terms. Therefore, the net contribution of $\rho$ and $\delta$ related exchange terms to $\sigma^{*}_{p\Delta^{++}}$ is larger than zero (gray dotted line).

\subsection{Isospin dependence of $\sigma^{*}_{N\Delta}(\sqrt{s},u,\alpha)$}\label{B2}

We also calculated the energy, density, and isospin asymmetry dependence of the individual $\sigma_{N \Delta}^{*}$, the results are shown in Fig. \ref{fig:graph3}. 
The values of individual $\sigma^{*}_{p\Delta}$ are shown in the top panels (a-d), while the individual $\sigma^{*}_{n\Delta}$ are represented in the bottom panels (e-h). 
The results for $\alpha$=0.3 with $u$ = 1.0, 2.0, and 3.0 are shown by blue solid, black dash-dot, and green dash-dot-dot lines, respectively. 
Similar to the results shown in Figs. \ref{fig.3a} and \ref{fig:cc33}, some individual cross sections are suppressed with increasing reduced densities and/or energies, especially at lower energies, while some elastic cross sections exhibit a slight enhancement with increasing densities at higher energies. A more detailed discussion of this energy dependence is addressed next and illustrated in Fig. \ref{fig.5}. 
In addition, the density dependence of $\sigma_{p \Delta^{++}}^{*}$ and $\sigma_{n \Delta^{-}}^{*}$ is stronger than that of other individual cross sections because the isospin matrices of these two channels are 3/2 for isospin scalar-vector terms and 9/4 for isospin vector-vector terms; there are larger values than those for other channels, as shown in Tab. \ref{table 1-1}. 

\begin{figure}[t]
    \centering
    \includegraphics[width=1\linewidth]{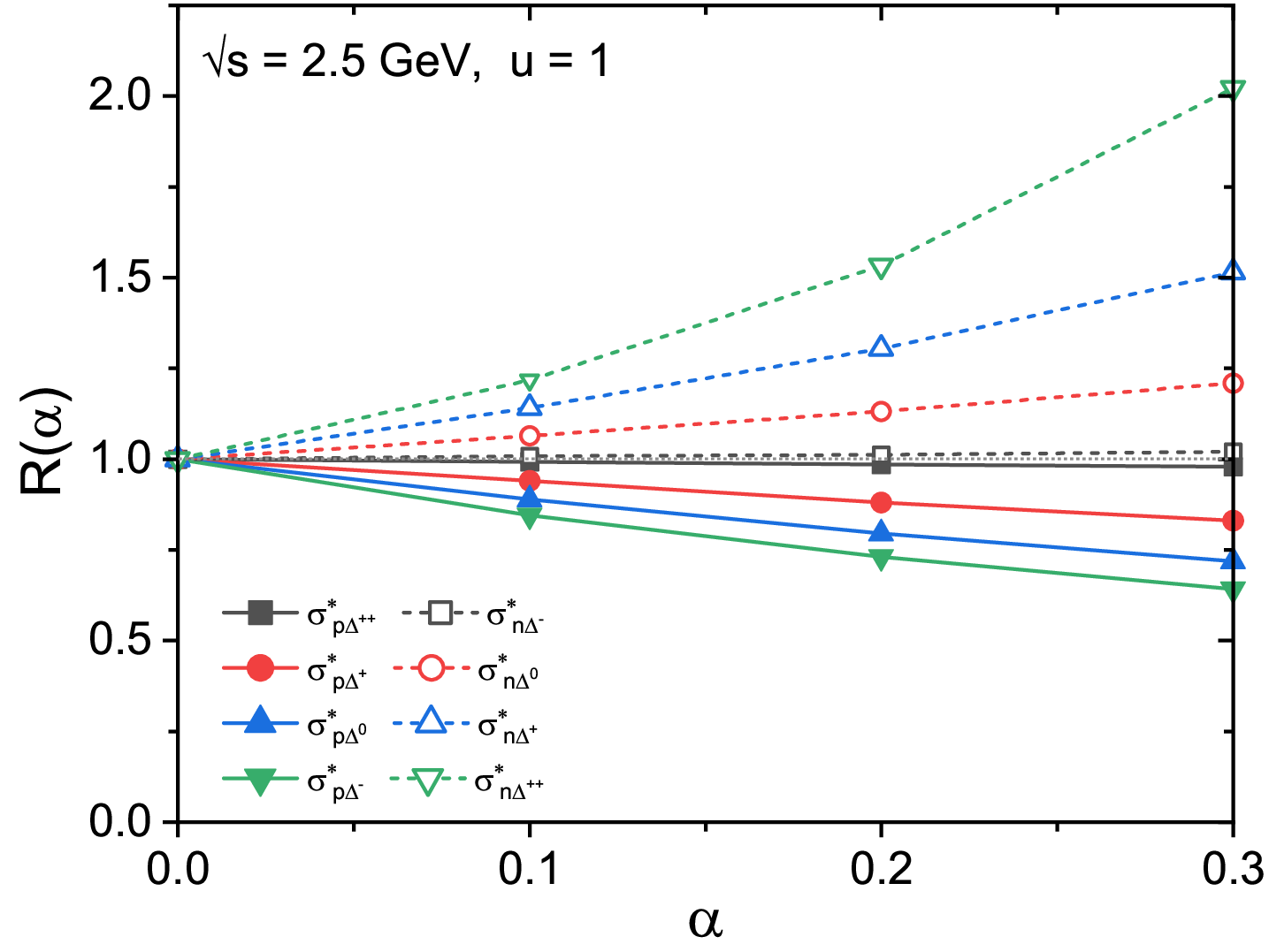}
    \caption{(Color online) $R(\alpha)=\sigma^*(\alpha)/\sigma^*(\alpha=0)$ ratios for all channels as a function of the isospin asymmetry $\alpha$ for $u$ = 1 at $\sqrt{s}$ = 2.5 GeV; the horizontal gray dotted line represents unity.}
    \label{fig.abc}
\end{figure}

\begin{figure}[t]
    \centering
    \includegraphics[width=1\linewidth]{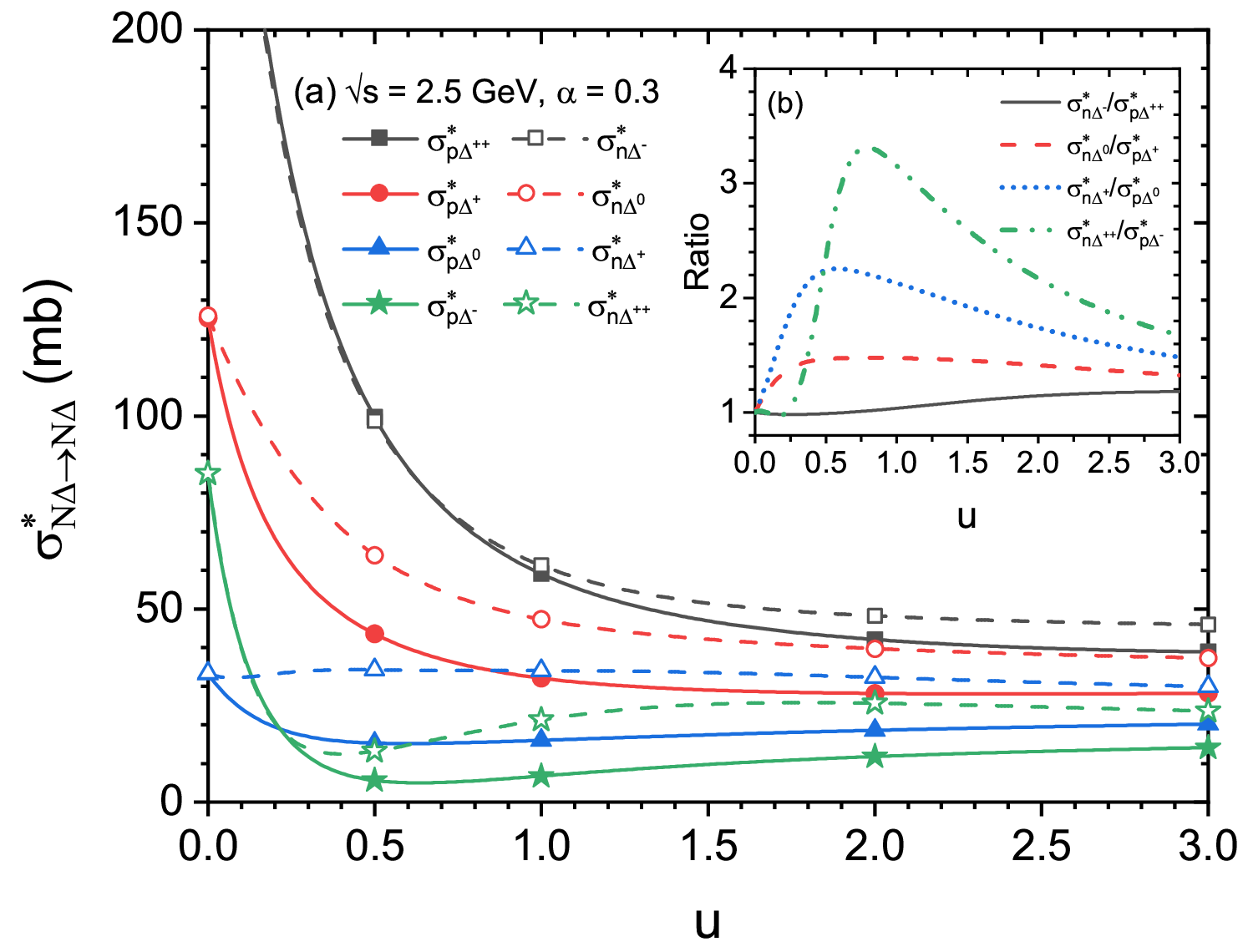}
   \caption{(Color online) Panel (a): individual $N\Delta$ elastic cross sections at $\sqrt{s}$= 2.5 GeV as a function of reduced density for asymmetric nuclear matter ($\alpha$=0.3). The inserted panel (b) shows the $\sigma^{*}_{n\Delta^{-}}$/$\sigma^{*}_{p\Delta^{++}}$, $\sigma^{*}_{n\Delta^{0}}$/$\sigma^{*}_{p\Delta^{+}}$, $\sigma^{*}_{n\Delta^{+}}$/$\sigma^{*}_{p\Delta^{0}}$, $\sigma^{*}_{n\Delta^{++}}$/$\sigma^{*}_{p\Delta^{-}}$ ratios.}
    \label{fig.5}
\end{figure}

In Fig. \ref{fig:graph3}, the calculated cross sections as a function of $\sqrt{s}$ for $\alpha=0.0$, $0.1$, and $0.3$ at $u=1$  are also represented by blue short dot, blue short dashed, and blue solid lines, respectively. 
As the isospin asymmetry $\alpha$ increases from 0.0 to 0.3, the individual cross sections of $p\Delta$ channels are suppressed, while the individual cross sections of $n\Delta$ channels are enhanced. 
Note that there is a clear splitting between individual $p\Delta$ and $n\Delta$ channels due to the different effective mass splitting effects on neutrons and protons, as well as differently charged $\Delta$ particles. 
This splitting effect is also observed in the $NN\rightarrow NN$, $NN\rightarrow N\Delta$, and $N\pi \rightarrow \Delta$ cross sections when the $\delta$ meson exchange is taken into account \cite{Li:2003vd,Li:2017pis,Li:2016xix}.  
In addition, the variation in the isospin-dependent elastic cross section of $n\Delta$ is relatively larger than that of $p\Delta$, while $\sigma^{*}_{p\Delta^{++}}$ (panel (a)) and $\sigma^{*}_{n\Delta^{-}}$ (panel (e)) show the weakest $\alpha$ dependence among these eight channels.

Therefore, there is an evident isospin dependence of $\sigma^{*}_{N\Delta\rightarrow N\Delta}$, as shown above. The isospin-dependent ratio $R(\alpha)=\sigma^*(\alpha)/\sigma^*(\alpha=0)$ for all channels for $u = 1$ at $\sqrt{s}$=2.5 GeV is depicted in Fig. \ref{fig.abc}. 
It can be found that the $R(\alpha)$ ratio deviates from unity (gray dotted line) and the ratio of $p\Delta$ channels (solid symbols) decreases while that for $n\Delta$ channels (open symbols) increases as $\alpha$ increases from 0.0 to 0.3. This results form the contribution of $\delta$ meson exchange to the effective masses of proton, neutron and $\Delta$-isobars, which exhibit opposite signs and further influence on the individual cross sections \cite{Li:2016xix}. 
Moreover, the ratio fulfill that $R(\alpha)_{n\Delta^{++}}>R(\alpha)_{n\Delta^{+}} > R(\alpha)_{n\Delta^{0}} > R(\alpha)_{n\Delta^{-}} > 1 > R(\alpha)_{p\Delta^{++}} > R(\alpha)_{p\Delta^{+}} > R(\alpha)_{p\Delta^{0}} > R(\alpha)_{p\Delta^{-}}$. 
At a higher isospin asymmetry parameter $\alpha = 0.3$, $R(\alpha)$=2.02, 1.02, 0.98, and 0.64 for $n\Delta^{++}$, $n\Delta^{-}$, $p\Delta^{++}$, and $p\Delta^{-}$ respectively. 
Comparing these ratios with those of $NN \rightarrow  N\Delta$ and $N\pi \rightarrow \Delta$ discussed in Refs. \cite{Li:2017pis,Li:2016xix}, it can be concluded that the isospin effect in $N\Delta\rightarrow N\Delta$ channels should not be negligible even at such a high $\sqrt{s}$.
Thus, concerning the yields of $\Delta$-isobars and corresponding daughter pions, the charged-pion ratio in intermediate-energy HICs should be influenced by the isospin-dependent in-medium correction on the $N\Delta$ elastic cross section, which comes from the isovector $\rho$ and $\delta$ meson fields \cite{Li:2016xix,Godbey:2021tbt}. 

As an extension to Fig. \ref{fig:graph3}, the individual $N\Delta$ cross sections for asymmetric nuclear matter ($\alpha=0.3$) at $\sqrt{s}$ = 2.5 GeV as a function of reduced density are shown in Fig. \ref{fig.5}(a). Note that the reduced density below approximately 0.5, $\sigma_{p\Delta^{++}(n\Delta^{-})}^{*}$ (black squares), $\sigma_{p\Delta^{+}(n\Delta^{0})}^{*}$ (red circles), and $\sigma_{p\Delta^{-}(n\Delta^{++})}^{*}$ (green stars) decrease rapidly with increasing density, while $\sigma_{p\Delta^{0}(n\Delta^{+})}^{*}$ (blue triangles) shows a much weaker dependence on the density. 
At density above approximately 1.0, the density dependence of all individual cross sections gradually weakens, eventually stabilizing.
These irregular density-dependent behaviours of the individual $N\Delta$ elastic cross sections can be explained by the integrated contributions of the spin and isospin matrices, density-dependent coupling constants, and density-dependent effective masses of nucleons and charged $\Delta$s.
Moreover, note that even within the range of 2-3 times normal density, there still exist evident differences between the channels which have the same isospin matrix parameter sets but different effective masses, such as an obvious splitting in $\sigma^{*}_{p\Delta^{++}}$ and $\sigma^{*}_{n\Delta^{-}}$.
This indicates that the influence of isovector $\rho$ and $\delta$ meson fields on the cross section is still visible at such a high density and can-not be ignored. 
In addition, the related  $\sigma^{*}_{n\Delta^{-}}$/$\sigma^{*}_{p\Delta^{++}}$,$\sigma^{*}_{n\Delta^{0}}$/$\sigma^{*}_{p\Delta^{+}}$, $\sigma^{*}_{n\Delta^{+}}$/$\sigma^{*}_{p\Delta^{0}}$, $\sigma^{*}_{n\Delta^{++}}$/$\sigma^{*}_{p\Delta^{-}}$ ratios are shown in the subgraph Fig. \ref{fig.5}(b). 
For each ratio, both the numerator and denominator share the same isospin matrix set, as shown in Tab. \ref{table 1-1}. 
The behaviors of these ratios with respect to density indicate that the influence of the splitting in the effective masses of nucleons and charged $\Delta$s on  $\sigma_{n\Delta^{++}}^{*}/\sigma_{p\Delta^{-}}^{*}$ (green dashed dot-dot line) is more pronounced than that of $\sigma_{n\Delta^{+}}^{*}/\sigma_{p\Delta^{0}}^{*}$, $\sigma_{n\Delta^{0}}^{*}/\sigma_{p\Delta^{+}}^{*}$, 
and $\sigma_{n\Delta^{-}}^{*}/\sigma_{p\Delta^{++}}^{*}$. 
It is understandable that the different coefficients of the contributions of the $\delta$ meson field in Eq. \ref{eq.7} to the effective mass of proton, neutron, $\Delta^{++}$, $\Delta^{+}$, $\Delta^{0}$, and $\Delta^{-}$ result in different effects on the splitting in the effective masses of the ingoing nucleon and $\Delta$, further influencing the individual cross sections as well as their ratios.

\subsection{Parametrization of $\sigma^{*}_{N\Delta}(\sqrt{s},u,\alpha)$}\label{B3}

\begin{figure}
    \centering
    \includegraphics[width=1\linewidth]{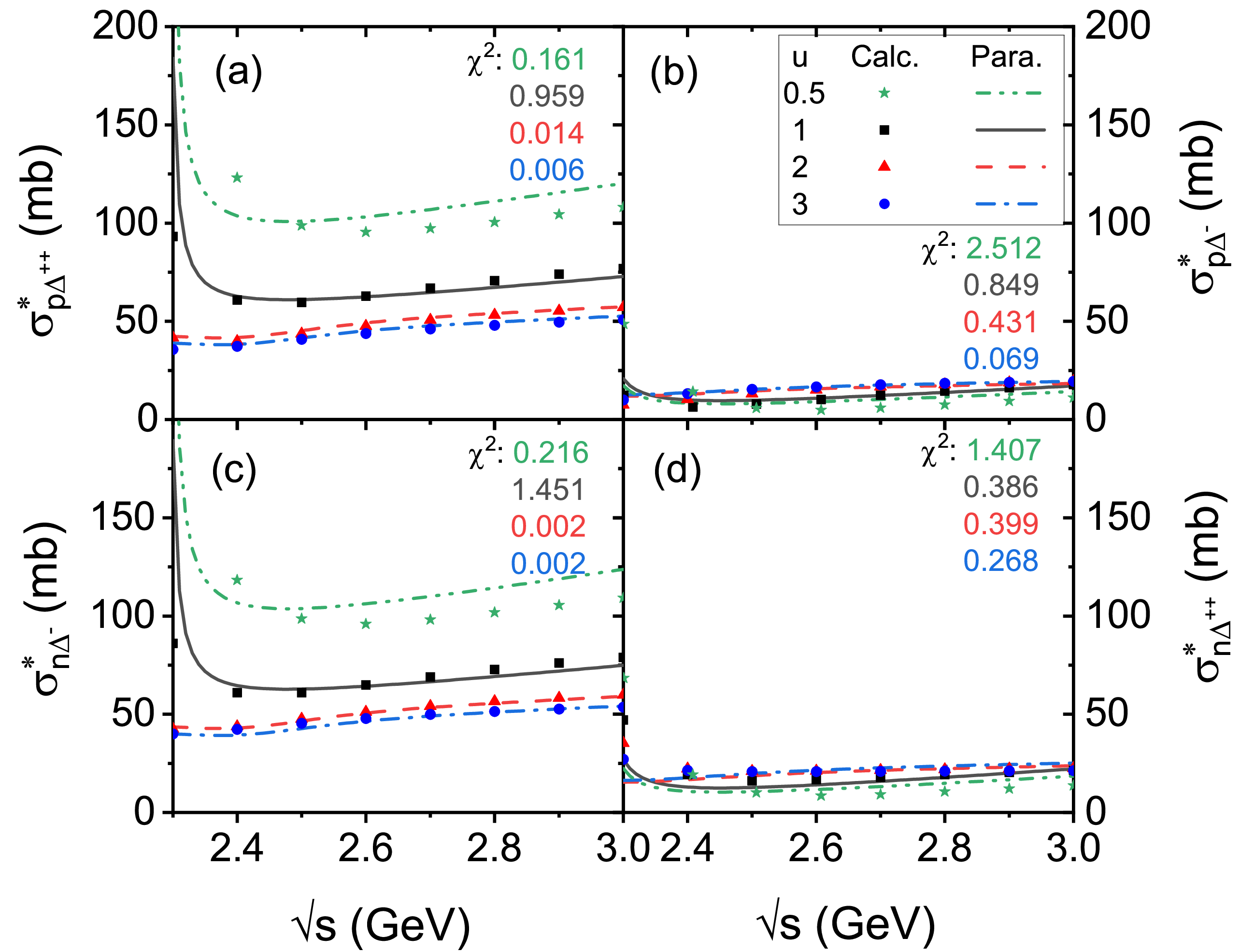}
   \caption{(Color online) 
  $\sigma^{*}_{p\Delta^{++}}$ (a), $\sigma^{*}_{p\Delta^{-}}$ (b), $\sigma^{*}_{n\Delta^{-}}$ (c), and $\sigma^{*}_{n\Delta^{++}}$ (d) as functions of the c.m. energy at $u$ = 0.5 (stars), 1 (squares), 2 (triangles) and 3 (circles) for $\alpha=0.2$, respectively. 
   The calculation results based on the RBUU approach are represented by symbols, while the parametrization results from Eq.\ref{formul} are represented by lines.}
    \label{fig.6}
\end{figure}

Lastly, to accurately and efficiently describe the dynamic processes of HICs and understand the properties of dense nuclear matter, 
the two-body cross section should be treated carefully in microscopic transport models. However, owing to the complex nature of cross sections, parametrized formulas based on theoretical calculations are commonly used in relevant models, such as the density- and energy-dependent formula for the $NN$ cross section proposed in Ref.\cite{Cai:1998iv}. They are also used in the isospin-dependent quantum molecular dynamics model \cite{Su:2016adl}.
The parametrized $NN$ elastic and inelastic cross sections are used in the Giessen Boltzmann–Uehling–Uhlenbeck \cite{Buss:2011mx} and relativistic Vlasov-Uehling-Uhlenbeck
\cite{Godbey:2021tbt}. 
Here, based on the above calculations within the RBUU theoretical framework, we propose the following parametrization for the energy ($\sqrt{s}$)-, density ($u$)-, and isospin ($\alpha$)-dependent $N\Delta$ elastic cross sections:
\begin{equation}
\begin{aligned}
   \sigma^{*}_{N\Delta \rightarrow N\Delta}(\sqrt{s}, u, \alpha) =
   &
   \left[a \left(\sqrt{s} + b\right) + \frac{c \left(\sqrt{s} + d\right)}{e + \left(\sqrt{s} + f\right)^2}\right] \\
   &\times 
   \left[g + h u + i \exp(j u)\right]  \\
   &\times 
   \left(1 + k \alpha + l \alpha^2\right),
\end{aligned}
\label{formul}
\end{equation}
where $a$ to $l$ are the adjustable parameters for each individual channel, and $\sqrt{s}$ is expressed in the unit of GeV. The energy, density, and isospin dependence of $\sigma^{*}_{N\Delta}$ can be expressed by the three parts of the parametrization in the brackets in sequence, respectively.
Figure \ref{fig.6} shows a comparison of the theoretically calculated results (symbols) with the parametrization results (lines) of $\sigma^{*}_{p\Delta^{++}}$, $\sigma^{*}_{p\Delta^{-}}$, $\sigma^{*}_{n\Delta^{-}}$, and $\sigma^{*}_{n\Delta^{++}}$ at $u$=0.5, 1, 2, and 3 for $\alpha$=0.2, respectively; the $\chi^2$ values are shown in each panel with corresponding colours, and the adjustable parameter sets of Eq. \ref{formul} for these four channels are shown in Table \ref{table-3}.
It can be seen that the parametrization can properly reproduce the microscopic calculation results within the c.m. energy region of $2.3 \leq \sqrt{s} \leq 3$ GeV and density range of $0.5 \leq u \leq 3$ at $\alpha$=0.2, which indicates that the proposed formula provides a reliable description of cross section within a wide range of energy, density, and isospin asymmetry, serving as a trustworthy input for transport model simulations of HICs.

\section{SUMMARY and Outlook}\label{sec4}

In this study, we calculated the energy-, density- and isospin-dependent $N\Delta \rightarrow N\Delta$ elastic cross sections $\sigma^{*}_{N\Delta}$ based on the density-dependent relativistic hadron field theory within the RBUU theoretical framework, in which the isovector $\delta$ meson field is further considered. 
The calculation results show that the decay width of $\Delta$ has a significant influence on $\sigma^{*}_{N\Delta}$ at lower energies ($\sqrt{s}\lesssim$ 2.2 GeV), while at higher energies ($\sqrt{s}\gtrsim$~2.2 GeV), the cross section becomes weakly dependent of the $\Delta$ resonance decay width. 
The total $\sigma^{*}_{N\Delta}$ is suppressed with increasing reduced density, and both $\rho$ and $\delta$ meson related exchange terms have non-negligible contributions to $\sigma^{*}_{N\Delta}$.  
By further analyzing the contribution of each $\rho$ and $\delta$ meson related exchange term to $\sigma^{*}_{p\Delta^{++}}$, we found that there exists a significant cancellation effect among these meson exchange terms due to the delicate balance of isovector $\rho$ and $\delta$ meson related terms.
Importantly, by including the $\delta$ meson field, individual $\sigma^{*}_{N\Delta}$ exhibit an evident isospin asymmetry dependence mainly due to the splitting in the effective masses of nucleons and $\Delta$ particles. 
As $\alpha$ increases, the $p\Delta$ related cross sections are suppressed, while the $n\Delta$ related cross sections are enhanced. 
Furthermore, the isospin effect introduced by the isovector $\rho$ and $\delta$ meson fields exerts an unignorable effect on the individual $N\Delta$ elastic cross sections even at 2-3 $\rho_{0}$.
In addition, a reliable parametrized formula for the energy-, density-, and isospin-dependent $N\Delta$ cross sections is proposed.

In the near future, the impact of the canonical momentum correction and threshold effect will be uniformly considered, and the parametrized formula for the $N\Delta \rightarrow N\Delta$ cross section will be improved and introduced into the UrQMD model to obtain a more comprehensive understanding of the production, evolution, and decay of $\Delta$ particles in heavy-ion collisions at intermediate energies, given that they are critical for making more reliable constraints on the high-density nuclear equation of state using pion related observables from HICs.


\section*{acknowledgements}
The work is supported in part by the National Natural Science Foundation of China (No. 12335008), the National Key Research and Development Program of China under (Grant No. 2023YFA1606402), the Zhejiang Provincial Natural Science Foundation of China (No. LQN25A050003), the Huzhou Natural Science Foundation (No. 2024YZ28), a project supported by the Scientific Research Fund of the Zhejiang Provincial Education Department (No. Y202353782), the Foundation of National Key Laboratory of Plasma Physics (Grant No. 6142A04230203). The authors are grateful to the C3S2 computing center in Huzhou University for calculation support.

\begin{widetext}
\begin{appendices}

\section{Appendix A}\label{App_A}
In this appendix, the analytical expressions of the collision terms of $\Delta$'s RBUU equation are presented. For the
spin matrices of $N_{2}\Delta \rightarrow N_{4}\Delta_{3}$ scattering, the scalar-scalar $D_{11}$, vector-vector $D_{12}$, and scalar-vector $D_{13}$ meson exchange components in differential cross sections are expressed as
\begin{equation}
\begin{aligned}
D_{11} = &\frac{4 g_{\Delta a}g_{\Delta b} g_{a} g_{b}}{9 m_{\Delta }^{* 2} m_{3 \Delta }^{* 2} (t-m_{a }^2)(t-m_{b }^2)} (-m_{2}^{* 2}-2 m_4^{* } m_2^{* }-m_4^{* 2}+t) (-m_{\Delta }^{* 2}-2 m_{3 \Delta }^{* } m_{\Delta }^{* }-m_{3 \Delta}^{* 2}+t) \\
& \left\{ m_{\Delta }^{* 4}+2 m_{3 \Delta }^{* }  m_{\Delta }^{* 3}-2 m_{\Delta }^{* 2} (t-6 m_{3 \Delta}^{* 2})+m_{\Delta }^{* } (2 m_{3\Delta }^{* 3}-2 t m_{3 \Delta }^{*})+(t-m_{3 \Delta }^{* 2})^2 \right\},
\end{aligned}
\end{equation}
\begin{equation}
\begin{aligned}
D_{12}= &\frac{8 g_{\Delta a}g_{\Delta b} g_{a} g_{b}}{9 m_{\Delta }^{* 2} m_{3 \Delta }^{* 2} (t-m_{a }^2)(t-m_{b }^2)} \bigg(  2 m_{3 \Delta }^{* 2} m_{\Delta }^{* 6} - 2 s m_{\Delta }^{* 6} - t m_{\Delta }^{* 6} + 4 m_{3 \Delta }^{* 3} m_{\Delta }^{* 5} - 4 s m_{3 \Delta }^{ *} m_{\Delta }^{* 5} + \\
& 2 t m_{3 \Delta }^{ *} m_{\Delta }^{* 5} + 24 m_{3 \Delta }^{* 4} m_{\Delta }^{* 4} + 2 s^2 m_{\Delta }^{* 4} + 3 t^2 m_{\Delta }^{* 4} - 26 s m_{3 \Delta }^{* 2} m_{\Delta }^{* 4} - 13 t m_{3 \Delta }^{* 2} m_{\Delta }^{* 4} + 6 s t m_{\Delta }^{* 4} + \\
& 4 m_{3 \Delta }^{* 5} m_{\Delta }^{* 3} - 8 s m_{3 \Delta }^{* 3} m_{\Delta }^{* 3} + 16 t m_{3 \Delta }^{* 3} m_{\Delta }^{* 3} + 4 m_4^{*4} m_{3 \Delta }^{ *} m_{\Delta }^{* 3} + 4 s^2 m_{3 \Delta }^{ *} m_{\Delta }^{* 3} - \\
& 4 t^2 m_{3 \Delta }^{ *} m_{\Delta }^{* 3} + 4 s t m_{3 \Delta }^{*} m_{\Delta }^{* 3} + 2 m_{3 \Delta }^{* 6} m_{\Delta }^{* 2} - 26 s m_{3 \Delta }^{* 4} m_{\Delta }^{* 2} - 13 t m_{3 \Delta }^{* 4} m_{\Delta }^{* 2} - \\
& 3 t^3 m_{\Delta }^{* 2} - 6 s t^2 m_{\Delta }^{* 2} + 24 s^2 m_{3 \Delta }^{* 2} m_{\Delta }^{* 2} + 14 t^2 m_{3 \Delta }^{* 2} m_{\Delta }^{* 2} + 32 s t m_{3 \Delta }^{* 2} m_{\Delta }^{* 2} - 4 s^2 t m_{\Delta }^{* 2} - \\
& 4 s m_{3 \Delta }^{* 5} m_{\Delta}^{*} + 2 t m_{3 \Delta }^{* 5} m_{\Delta}^{*} + 4 m_2^{*4} m_{3 \Delta }^{* 3} m_{\Delta}^{*} + 4 s^2 m_{3 \Delta }^{* 3} m_{\Delta}^{*} - 4 t^2 m_{3 \Delta }^{* 3} m_{\Delta}^{*} + \\
& 4 s t m_{3 \Delta }^{* 3} m_{\Delta}^{*} + 2 t^3 m_{3 \Delta }^{ *} m_{\Delta}^{*} - 2 s m_{3 \Delta }^{* 6} - t m_{3 \Delta }^{* 6} + t^4 + 2 s^2 m_{3 \Delta }^{* 4} + 3 t^2 m_{3 \Delta }^{* 4} + \\
& 6 s t m_{3 \Delta }^{* 4} + 2 s t^3 + 2 s^2 t^2 - 3 t^3 m_{3 \Delta }^{* 2} - 6 s t^2 m_{3 \Delta }^{* 2} - 4 s^2 t m_{3 \Delta }^{* 2} + \\
& 2 m_{2}^{*} m_{4}^{*} \left( -m_{\Delta }^{* 6} + 4 m_{3 \Delta }^{ *} m_{\Delta }^{* 5} + \left( 3 t - 5 m_{3 \Delta }^{* 2} \right) m_{\Delta }^{* 4} - 8 m_{3 \Delta }^{ *} \left( t - 5 m_{3 \Delta }^{* 2} \right) m_{\Delta }^{* 3} + \right. \\
& \left. \left( -5 m_{3 \Delta }^{* 4} + 8 t m_{3 \Delta }^{* 2} - 3 t^2 \right) m_{\Delta }^{* 2} + 4 m_{3 \Delta }^{ *} \left( t - m_{3 \Delta }^{* 2} \right)^2 m_{\Delta}^{*} + \left( t - m_{3 \Delta }^{* 2} \right)^3 \right) + \\
& m_4^{*2} \left( m_{\Delta }^{* 6} + 2 m_{3 \Delta }^{ *} m_{\Delta }^{* 5} + \left( 9 m_{3 \Delta }^{* 2} - 2 s - 3 t \right) m_{\Delta }^{* 4} - 4 \left( 5 m_{3 \Delta }^{* 3} + 2 s m_{3 \Delta }^{ *} \right) m_{\Delta }^{* 3} + \right. \\
& \left. \left( 9 m_{3 \Delta }^{* 4} - 12 (2 s + t) m_{3 \Delta }^{* 2} + t (4 s + 3 t) \right) m_{\Delta }^{* 2} - 2 m_{3 \Delta }^{ *} \left( t - m_{3 \Delta }^{* 2} \right)^2 m_{\Delta}^{*} - \right. \\
& \left. \left( t - m_{3 \Delta }^{* 2} \right)^2 \left( -m_{3 \Delta }^{* 2} + 2 s + t \right) \right) + \\
& m_2^{*2} \left( m_{\Delta }^{* 6} - 2 m_{3 \Delta }^{ *} m_{\Delta }^{* 5} + \left( 9 m_{3 \Delta }^{* 2} - 2 s - 3 t \right) m_{\Delta }^{* 4} + 4 m_{3 \Delta }^{ *} \left( t - 5 m_{3 \Delta }^{* 2} \right) m_{\Delta }^{* 3} + \right. \\
& \left. \left( 9 m_{3 \Delta }^{* 4} - 12 (2 s + t) m_{3 \Delta }^{* 2} + t (4 s + 3 t) \right) m_{\Delta }^{* 2} + 2 m_{3 \Delta }^{ *} \left( m_{3 \Delta }^{* 4} - 4 s m_{3 \Delta }^{* 2} - t^2 \right) m_{\Delta}^{*} - \right. \\
& \left. \left( t - m_{3 \Delta }^{* 2} \right)^2 \left( -m_{3 \Delta }^{* 2} + 2 s + t \right) + 2 m_4^{*2} \left( m_{\Delta }^{* 4} - 2 \left( t - 6 m_{3 \Delta }^{* 2} \right) m_{\Delta }^{* 2} + \left( t - m_{3 \Delta }^{* 2} \right)^2 \right) \right) \bigg),
\end{aligned}
\end{equation}
\begin{equation}
\begin{aligned}
D_{13}= & \frac{16 g_{\Delta a} g_{\Delta b} g_{a} g_{b}}{9 m_{\Delta}^{*2} m_{3 \Delta}^{*2} (t-m_{a }^2)(t-m_{b }^2)} \left\{m_{\Delta}^{*4} + 2 m_{3 \Delta }^{*} m_{\Delta}^{*3} - 2 m_{\Delta}^{*2} (t - 6 m_{3 \Delta}^{*2}) + m_{\Delta }^{*} (2 m_{3 \Delta}^{*3} - 2 t m_{3 \Delta }^{*}) + \right.\\
&\left. (t - m_{3 \Delta}^{*2})^2\right\} \left\{m_2^{*3} m_{3 \Delta }^{*} + m_{4}^{*} m_2^{*2} m_{3 \Delta }^{*} +
m_{2}^{*} (m_{\Delta }^{*} (m_{3 \Delta}^{*2} + m_4^{*2} - s) + m_{3 \Delta }^{*} (m_{3 \Delta}^{*2} - s - t)) + \right.\\
&\left. m_{4}^{*} (m_{\Delta}^{*3} + m_{3 \Delta }^{*} m_{\Delta}^{*2} - s m_{3 \Delta }^{*} - m_{\Delta }^{*} (-m_4^{*2} + s + t))\right\}.
\end{aligned}
\end{equation}
Correspondingly, $D_{1-10}$ can be expressed as follows:
\begin{equation}
    \begin{aligned}
&D_{1}=D_{11}\left(m_{a} \rightarrow m_{\sigma},m_{b} \rightarrow m_{\sigma},g_{a} \rightarrow g_{\sigma},g_{b} \rightarrow g_{\sigma},g_{\Delta a} \rightarrow g_{\Delta\sigma},g_{\Delta  b} \rightarrow g_{\Delta\sigma}\right),&\\
&D_{2}=D_{12}\left(m_{a} \rightarrow m_{\omega},m_{b} \rightarrow m_{\omega},g_{a} \rightarrow g_{\omega},g_{b} \rightarrow g_{\omega},g_{\Delta a} \rightarrow g_{\Delta\omega},g_{\Delta  b} \rightarrow g_{\Delta\omega}\right),&\\
&D_{3}=D_{13}\left(m_{a} \rightarrow m_{\sigma},m_{b} \rightarrow m_{\omega},g_{a} \rightarrow g_{\sigma},g_{b} \rightarrow g_{\omega},g_{\Delta a} \rightarrow g_{\Delta\sigma},g_{\Delta  b} \rightarrow g_{\Delta\omega}\right),&\\
&D_{4}=D_{12}\left(m_{a} \rightarrow m_{\rho},m_{b} \rightarrow m_{\rho},g_{a} \rightarrow g_{\rho},g_{b} \rightarrow g_{\rho},g_{\Delta a} \rightarrow g_{\Delta\rho},g_{\Delta  b} \rightarrow g_{\Delta \rho}\right),&\\
&D_{5}=D_{11}\left(m_{a} \rightarrow m_{\delta},m_{b} \rightarrow m_{\delta},g_{a} \rightarrow g_{\delta},g_{b} \rightarrow g_{\delta},g_{\Delta a} \rightarrow g_{\Delta \delta},g_{\Delta  b} \rightarrow g_{\Delta \delta}\right),&\\
&D_{6}=D_{13}\left(m_{a} \rightarrow m_{\delta},m_{b} \rightarrow m_{\rho},g_{a} \rightarrow g_{\delta},g_{b} \rightarrow g_{\rho},g_{\Delta a} \rightarrow g_{\Delta\delta},g_{\Delta  b} \rightarrow g_{\Delta \rho}\right),&\\
&D_{7}= 2 D_{11}\left(m_{a} \rightarrow m_{\sigma},m_{b} \rightarrow m_{\delta},g_{a} \rightarrow g_{\sigma},g_{b} \rightarrow g_{\delta},g_{\Delta a} \rightarrow g_{\Delta\sigma},g_{\Delta  b} \rightarrow g_{\Delta\delta}\right),&\\
&D_{8}=D_{13}\left(m_{a} \rightarrow m_{\sigma},m_{b} \rightarrow m_{\rho},g_{a} \rightarrow g_{\sigma},g_{b} \rightarrow g_{\rho},g_{\Delta a} \rightarrow g_{\Delta\sigma},g_{\Delta  b} \rightarrow g_{\Delta\rho}\right),&\\
&D_{9}=D_{13}\left(m_{a} \rightarrow m_{\omega},m_{b} \rightarrow m_{\delta},g_{a} \rightarrow g_{\omega},g_{b} \rightarrow g_{\delta},g_{\Delta a} \rightarrow g_{\Delta\omega},g_{\Delta  b} \rightarrow g_{\Delta\delta}\right),&\\
&D_{10}=2 D_{12}\left(m_{a} \rightarrow m_{\omega},m_{b} \rightarrow m_{\rho},g_{a} \rightarrow g_{\omega},g_{b} \rightarrow g_{\rho},g_{\Delta a} \rightarrow g_{\Delta\omega},g_{\Delta  b} \rightarrow g_{\Delta\rho}\right),&
\end{aligned}
\end{equation}
where $s$, $t$, and $u$ are Mandelstam variables  defined as
\begin{equation}
    \begin{aligned}
& s=\left(p_{1}+p_{2}\right)^{2}=\left[E_{\Delta}^{*}(p)+E^{*}\left(p_{2}\right)\right]^{2}-\left(\mathbf{p}+\mathbf{p}_{2}\right)^{2},\\
& t = m^{* 2}_{\Delta} +m^{* 2}_{3\Delta} - \left(\frac{(m^{* 2}_{\Delta} - m^{* 2}_{2} + s)
(m^{* 2}_{3\Delta} - m^{* 2}_{4} + s)}{2s}\right) + 2\left|\mathbf{p} \| \mathbf{p}_{3}\right| \cos \theta,\\
& u= m_{\Delta}^{* 2}+ m_{2}^{* 2}+ m_{3\Delta}^{* 2}+m_{4}^{* 2}-s-t.\\
\end{aligned}
\end{equation}
The scattering angle in the c.m. system is $\theta$, and
\begin{equation}
\begin{aligned}
&\left|\mathbf{p}\right|=\frac{1}{2 \sqrt{s}} \sqrt{\left(s-m_2^{* 2}-m_{\Delta}^{* 2}\right)^{2}-4 m_2^{* 2} m_{\Delta}^{* 2}},\\
& \left|\mathbf{p}_{3}\right|=\frac{1}{2 \sqrt{s}} \sqrt{\left(s-m_{3 \Delta}^{* 2}-m_{4 }^{* 2}\right)^{2}-4 m_{3 \Delta}^{* 2} m_{4}^{* 2}}.
\end{aligned}
\end{equation}

\newpage
\section{Appendix B}\label{App_B}

\begin{table}[htbp!]
\renewcommand{\arraystretch}{1.5} 
\centering
    \caption{Parameter sets of Eq.\ref{formul} for the in-medium energy-, density-, and isospin-dependent $\sigma^{*}_{p\Delta^{++}}$, $\sigma^{*}_{n\Delta^{-}}$, $\sigma^{*}_{p\Delta^{-}}$, and $\sigma^{*}_{n\Delta^{++}}$ elastic channels within the ranges of 2.3 $\leq \sqrt{s} \leq $ 3 GeV and $0.5 \leq u \leq 3$ for $\alpha$=0.2. Here, $\sqrt{s}$ is the c.m. energy, $u= \rho/\rho_{0}$ is the reduced density, and $\alpha$ is the isospin asymmetry degree.}
 \setlength{\tabcolsep}{2mm}{
\begin{tabular}{ccccccccc}
\hline\hline
& \multicolumn{2}{c}{$\sigma_{ p\Delta^{++}}^{*}$}  & \multicolumn{2}{c}{$\sigma_{ n\Delta^{-}}^{*}$} & \multicolumn{2}{c}{$\sigma_{ p\Delta^{-}}^{*}$} & \multicolumn{2}{c}{$\sigma_{ n\Delta^{++}}^{*}$} \\ \hline
& $0.5 \leq u < 1.5$& $1.5\leq u \leq 3$ &$0.5 \leq u < 1.5$ & $1.5\leq u \leq 3$ & $0.5 \leq u < 1.5$& $1.5\leq u \leq 3$ & $0.5 \leq u < 1.5$ & $1.5\leq u \leq 3$\\
$a$  & 3.591  & 2.629 & 3.591 & 2.629&2.419 & 1.0951 &2.419 & 1.0951   \\
$b$  & -0.696 & 1.209 & -0.696 & 1.209 &-2.103& 1.366 &-2.103 & 1.366   \\
$c$  & 1.466   & 0.00354  & 1.466 & 0.00354&0.0310 & 0.0311 &0.0310 & 0.0311  \\
$d$  & -0.394  & -14.547 & -0.394 & -14.547&3.489 & -3.822 & 3.489& -3.822   \\
$e$  & -123.877  & 0.0283  & -123.877 &0.0283 &-1.070 & 0.0495 &-1.070 &0.0495    \\
$f$  & 8.837  & -2.399 & 8.837 &-2.399 &-1.227 & -2.323 &-1.227 & -2.323   \\
$g$  & 6.923  & 5.139 & 6.923 & 5.139 & 2.441 &0.821 & 2.441 &0.821  \\
$h$  & -0.572  & -0.116 &-0.572 & -0.116 & 2.83 &0.2502 & 2.83 &0.2502  \\
$i$  & 25.509  & 9.450 & 25.509 & 9.450 &3.160 &2.9902  & 3.160 &2.9902\\
$j$  & -2.393 & -1.575 & -2.393 & -1.575 &-0.00429 &0.00113 &-0.00429 & 0.00113   \\
$k$  & -0.0867  & -0.0867 & 0.060 & 0.0607 &  -0.378  & -0.378 & 0.430 & 0.430\\
$l$  & 0.0552  & 0.0552 & 0.0189 &0.0189  &  -0.652 &  -0.652 & 1.793 &  1.793\\ 
\hline\hline
\end{tabular}}
\label{table-3}
\end{table}

\newpage
	
\end{appendices}

\end{widetext}

\bibliography{ref.bib}

\begin{thebibliography}{71}%
\makeatletter
\providecommand \@ifxundefined [1]{%
 \@ifx{#1\undefined}
}%
\providecommand \@ifnum [1]{%
 \ifnum #1\expandafter \@firstoftwo
 \else \expandafter \@secondoftwo
 \fi
}%
\providecommand \@ifx [1]{%
 \ifx #1\expandafter \@firstoftwo
 \else \expandafter \@secondoftwo
 \fi
}%
\providecommand \natexlab [1]{#1}%
\providecommand \enquote  [1]{``#1''}%
\providecommand \bibnamefont  [1]{#1}%
\providecommand \bibfnamefont [1]{#1}%
\providecommand \citenamefont [1]{#1}%
\providecommand \href@noop [0]{\@secondoftwo}%
\providecommand \href [0]{\begingroup \@sanitize@url \@href}%
\providecommand \@href[1]{\@@startlink{#1}\@@href}%
\providecommand \@@href[1]{\endgroup#1\@@endlink}%
\providecommand \@sanitize@url [0]{\catcode `\\12\catcode `\$12\catcode
  `\&12\catcode `\#12\catcode `\^12\catcode `\_12\catcode `\%12\relax}%
\providecommand \@@startlink[1]{}%
\providecommand \@@endlink[0]{}%
\providecommand \url  [0]{\begingroup\@sanitize@url \@url }%
\providecommand \@url [1]{\endgroup\@href {#1}{\urlprefix }}%
\providecommand \urlprefix  [0]{URL }%
\providecommand \Eprint [0]{\href }%
\providecommand \doibase [0]{http://dx.doi.org/}%
\providecommand \selectlanguage [0]{\@gobble}%
\providecommand \bibinfo  [0]{\@secondoftwo}%
\providecommand \bibfield  [0]{\@secondoftwo}%
\providecommand \translation [1]{[#1]}%
\providecommand \BibitemOpen [0]{}%
\providecommand \bibitemStop [0]{}%
\providecommand \bibitemNoStop [0]{.\EOS\space}%
\providecommand \EOS [0]{\spacefactor3000\relax}%
\providecommand \BibitemShut  [1]{\csname bibitem#1\endcsname}%
\let\auto@bib@innerbib\@empty
\bibitem [{\citenamefont {Danielewicz}\ \emph {et~al.}(2002)\citenamefont
  {Danielewicz}, \citenamefont {Lacey},\ and\ \citenamefont
  {Lynch}}]{Danielewicz:2002pu}%
  \BibitemOpen
  \bibfield  {author} {\bibinfo {author} {\bibfnamefont {P.}~\bibnamefont
  {Danielewicz}}, \bibinfo {author} {\bibfnamefont {R.}~\bibnamefont {Lacey}},
  \ and\ \bibinfo {author} {\bibfnamefont {W.~G.}\ \bibnamefont {Lynch}},\
  }\href {\doibase 10.1126/science.1078070} {\bibfield  {journal} {\bibinfo
  {journal} {Science}\ }\textbf {\bibinfo {volume} {298}},\ \bibinfo {pages}
  {1592} (\bibinfo {year} {2002})}\BibitemShut {NoStop}%
\bibitem [{\citenamefont {Li}\ \emph {et~al.}(2008)\citenamefont {Li},
  \citenamefont {Chen},\ and\ \citenamefont {Ko}}]{Li:2008gp}%
  \BibitemOpen
  \bibfield  {author} {\bibinfo {author} {\bibfnamefont {B.~A.}\ \bibnamefont
  {Li}}, \bibinfo {author} {\bibfnamefont {L.~W.}\ \bibnamefont {Chen}}, \ and\
  \bibinfo {author} {\bibfnamefont {C.~M.}\ \bibnamefont {Ko}},\ }\href
  {\doibase 10.1016/j.physrep.2008.04.005} {\bibfield  {journal} {\bibinfo
  {journal} {Phys. Rept.}\ }\textbf {\bibinfo {volume} {464}},\ \bibinfo
  {pages} {113} (\bibinfo {year} {2008})}\BibitemShut {NoStop}%
\bibitem [{\citenamefont {Sorensen}\ \emph {et~al.}(2024)\citenamefont
  {Sorensen}, \citenamefont {Agarwal}, \citenamefont {W.~Brown} \emph
  {et~al.}}]{Sorensen:2023zkk}%
  \BibitemOpen
  \bibfield  {author} {\bibinfo {author} {\bibfnamefont {A.}~\bibnamefont
  {Sorensen}}, \bibinfo {author} {\bibfnamefont {K.}~\bibnamefont {Agarwal}},
  \bibinfo {author} {\bibfnamefont {K.}~\bibnamefont {W.~Brown}},  \emph
  {et~al.},\ }\href {\doibase 10.1016/j.ppnp.2023.104080} {\bibfield  {journal}
  {\bibinfo  {journal} {Prog. Part. Nucl. Phys.}\ }\textbf {\bibinfo {volume}
  {134}},\ \bibinfo {pages} {104080} (\bibinfo {year} {2024})}\BibitemShut
  {NoStop}%
\bibitem [{\citenamefont {Li}\ \emph {et~al.}(2021)\citenamefont {Li},
  \citenamefont {Cai}, \citenamefont {Xie},\ and\ \citenamefont
  {Zhang}}]{Li:2021thg}%
  \BibitemOpen
  \bibfield  {author} {\bibinfo {author} {\bibfnamefont {B.~A.}\ \bibnamefont
  {Li}}, \bibinfo {author} {\bibfnamefont {B.~J.}\ \bibnamefont {Cai}},
  \bibinfo {author} {\bibfnamefont {W.~J.}\ \bibnamefont {Xie}}, \ and\
  \bibinfo {author} {\bibfnamefont {N.~B.}\ \bibnamefont {Zhang}},\ }\href
  {\doibase 10.3390/universe7060182} {\bibfield  {journal} {\bibinfo  {journal}
  {Universe}\ }\textbf {\bibinfo {volume} {7}},\ \bibinfo {pages} {182}
  (\bibinfo {year} {2021})}\BibitemShut {NoStop}%
\bibitem [{\citenamefont {Stoecker}\ and\ \citenamefont
  {Greiner}(1986)}]{Stoecker:1986ci}%
  \BibitemOpen
  \bibfield  {author} {\bibinfo {author} {\bibfnamefont {H.}~\bibnamefont
  {Stoecker}}\ and\ \bibinfo {author} {\bibfnamefont {W.}~\bibnamefont
  {Greiner}},\ }\href {\doibase 10.1016/0370-1573(86)90131-6} {\bibfield
  {journal} {\bibinfo  {journal} {Phys. Rept.}\ }\textbf {\bibinfo {volume}
  {137}},\ \bibinfo {pages} {277} (\bibinfo {year} {1986})}\BibitemShut
  {NoStop}%
\bibitem [{\citenamefont {Lattimer}\ and\ \citenamefont
  {Prakash}(2000)}]{Lattimer:2000kb}%
  \BibitemOpen
  \bibfield  {author} {\bibinfo {author} {\bibfnamefont {J.~M.}\ \bibnamefont
  {Lattimer}}\ and\ \bibinfo {author} {\bibfnamefont {M.}~\bibnamefont
  {Prakash}},\ }\href {\doibase 10.1016/S0370-1573(00)00019-3} {\bibfield
  {journal} {\bibinfo  {journal} {Phys. Rept.}\ }\textbf {\bibinfo {volume}
  {333}},\ \bibinfo {pages} {121} (\bibinfo {year} {2000})}\BibitemShut
  {NoStop}%
\bibitem [{\citenamefont {Vretenar}\ \emph {et~al.}(2005)\citenamefont
  {Vretenar}, \citenamefont {Afanasjev}, \citenamefont {Lalazissis},\ and\
  \citenamefont {Ring}}]{Vretenar:2005zz}%
  \BibitemOpen
  \bibfield  {author} {\bibinfo {author} {\bibfnamefont {D.}~\bibnamefont
  {Vretenar}}, \bibinfo {author} {\bibfnamefont {A.~V.}\ \bibnamefont
  {Afanasjev}}, \bibinfo {author} {\bibfnamefont {G.~A.}\ \bibnamefont
  {Lalazissis}}, \ and\ \bibinfo {author} {\bibfnamefont {P.}~\bibnamefont
  {Ring}},\ }\href {\doibase 10.1016/j.physrep.2004.10.001} {\bibfield
  {journal} {\bibinfo  {journal} {Phys. Rept.}\ }\textbf {\bibinfo {volume}
  {409}},\ \bibinfo {pages} {101} (\bibinfo {year} {2005})}\BibitemShut
  {NoStop}%
\bibitem [{\citenamefont {Baldo}\ and\ \citenamefont
  {Burgio}(2016)}]{Baldo:2016jhp}%
  \BibitemOpen
  \bibfield  {author} {\bibinfo {author} {\bibfnamefont {M.}~\bibnamefont
  {Baldo}}\ and\ \bibinfo {author} {\bibfnamefont {G.~F.}\ \bibnamefont
  {Burgio}},\ }\href {\doibase 10.1016/j.ppnp.2016.06.006} {\bibfield
  {journal} {\bibinfo  {journal} {Prog. Part. Nucl. Phys.}\ }\textbf {\bibinfo
  {volume} {91}},\ \bibinfo {pages} {203} (\bibinfo {year} {2016})}\BibitemShut
  {NoStop}%
\bibitem [{\citenamefont {Russotto}\ \emph {et~al.}(2016)\citenamefont
  {Russotto} \emph {et~al.}}]{Russotto:2016ucm}%
  \BibitemOpen
  \bibfield  {author} {\bibinfo {author} {\bibfnamefont {P.}~\bibnamefont
  {Russotto}} \emph {et~al.} (\bibinfo {collaboration} {ASY-EOS
  Collaboration}),\ }\href {\doibase 10.1103/PhysRevC.94.034608} {\bibfield
  {journal} {\bibinfo  {journal} {Phys. Rev. C}\ }\textbf {\bibinfo {volume}
  {94}},\ \bibinfo {pages} {034608} (\bibinfo {year} {2016})}\BibitemShut
  {NoStop}%
\bibitem [{\citenamefont {Huth}\ \emph {et~al.}(2022)\citenamefont {Huth},
  \citenamefont {Pang}, \citenamefont {Tews} \emph {et~al.}}]{Huth:2021bsp}%
  \BibitemOpen
  \bibfield  {author} {\bibinfo {author} {\bibfnamefont {S.}~\bibnamefont
  {Huth}}, \bibinfo {author} {\bibfnamefont {P.~T.~H.}\ \bibnamefont {Pang}},
  \bibinfo {author} {\bibfnamefont {I.}~\bibnamefont {Tews}},  \emph {et~al.},\
  }\href {\doibase 10.1038/s41586-022-04750-w} {\bibfield  {journal} {\bibinfo
  {journal} {Nature}\ }\textbf {\bibinfo {volume} {606}},\ \bibinfo {pages}
  {276} (\bibinfo {year} {2022})}\BibitemShut {NoStop}%
\bibitem [{\citenamefont {Zhou}\ and\ \citenamefont
  {Yang}(2022)}]{Zhou:2022pxl}%
  \BibitemOpen
  \bibfield  {author} {\bibinfo {author} {\bibfnamefont {X.~H.}\ \bibnamefont
  {Zhou}}\ and\ \bibinfo {author} {\bibfnamefont {J.~C.}\ \bibnamefont {Yang}}
  (\bibinfo {collaboration} {HIAF project Team}),\ }\href {\doibase
  10.1007/s43673-022-00064-1} {\bibfield  {journal} {\bibinfo  {journal} {AAPPS
  Bull.}\ }\textbf {\bibinfo {volume} {32}},\ \bibinfo {pages} {35} (\bibinfo
  {year} {2022})}\BibitemShut {NoStop}%
\bibitem [{\citenamefont {Ablyazimov}\ \emph {et~al.}(2017)\citenamefont
  {Ablyazimov} \emph {et~al.}}]{CBM:2016kpk}%
  \BibitemOpen
  \bibfield  {author} {\bibinfo {author} {\bibfnamefont {T.}~\bibnamefont
  {Ablyazimov}} \emph {et~al.} (\bibinfo {collaboration} {CBM Collaboration}),\
  }\href {\doibase 10.1140/epja/i2017-12248-y} {\bibfield  {journal} {\bibinfo
  {journal} {Eur. Phys. J. A}\ }\textbf {\bibinfo {volume} {53}},\ \bibinfo
  {pages} {60} (\bibinfo {year} {2017})}\BibitemShut {NoStop}%
\bibitem [{\citenamefont {Chen}\ \emph {et~al.}(2024)\citenamefont {Chen},
  \citenamefont {Dong}, \citenamefont {He} \emph {et~al.}}]{Chen:2024zwk}%
  \BibitemOpen
  \bibfield  {author} {\bibinfo {author} {\bibfnamefont {J.}~\bibnamefont
  {Chen}}, \bibinfo {author} {\bibfnamefont {X.}~\bibnamefont {Dong}}, \bibinfo
  {author} {\bibfnamefont {X.}~\bibnamefont {He}},  \emph {et~al.},\ }\href
  {\doibase 10.1007/s41365-024-01591-2} {\bibfield  {journal} {\bibinfo
  {journal} {Nucl. Sci. Tech.}\ }\textbf {\bibinfo {volume} {35}},\ \bibinfo
  {pages} {214} (\bibinfo {year} {2024})}\BibitemShut {NoStop}%
\bibitem [{\citenamefont {Abgaryan}\ \emph {et~al.}(2022)\citenamefont
  {Abgaryan} \emph {et~al.}}]{MPD:2022qhn}%
  \BibitemOpen
  \bibfield  {author} {\bibinfo {author} {\bibfnamefont {V.}~\bibnamefont
  {Abgaryan}} \emph {et~al.} (\bibinfo {collaboration} {MPD Collaboration}),\
  }\href {\doibase 10.1140/epja/s10050-022-00750-6} {\bibfield  {journal}
  {\bibinfo  {journal} {Eur. Phys. J. A}\ }\textbf {\bibinfo {volume} {58}},\
  \bibinfo {pages} {140} (\bibinfo {year} {2022})}\BibitemShut {NoStop}%
\bibitem [{\citenamefont {Li}\ \emph {et~al.}(2006)\citenamefont {Li},
  \citenamefont {Li}, \citenamefont {Soff}, \citenamefont {Bleicher},\ and\
  \citenamefont {Stoecker}}]{Li:2005gfa}%
  \BibitemOpen
  \bibfield  {author} {\bibinfo {author} {\bibfnamefont {Q.~F.}\ \bibnamefont
  {Li}}, \bibinfo {author} {\bibfnamefont {Z.~X.}\ \bibnamefont {Li}}, \bibinfo
  {author} {\bibfnamefont {S.}~\bibnamefont {Soff}}, \bibinfo {author}
  {\bibfnamefont {M.}~\bibnamefont {Bleicher}}, \ and\ \bibinfo {author}
  {\bibfnamefont {H.}~\bibnamefont {Stoecker}},\ }\href {\doibase
  10.1088/0954-3899/32/2/007} {\bibfield  {journal} {\bibinfo  {journal} {J.
  Phys. G}\ }\textbf {\bibinfo {volume} {32}},\ \bibinfo {pages} {151}
  (\bibinfo {year} {2006})}\BibitemShut {NoStop}%
\bibitem [{\citenamefont {Xiao}\ \emph {et~al.}(2009)\citenamefont {Xiao},
  \citenamefont {Li}, \citenamefont {Chen}, \citenamefont {Yong},\ and\
  \citenamefont {Zhang}}]{Xiao:2008vm}%
  \BibitemOpen
  \bibfield  {author} {\bibinfo {author} {\bibfnamefont {Z.~G.}\ \bibnamefont
  {Xiao}}, \bibinfo {author} {\bibfnamefont {B.~A.}\ \bibnamefont {Li}},
  \bibinfo {author} {\bibfnamefont {L.~W.}\ \bibnamefont {Chen}}, \bibinfo
  {author} {\bibfnamefont {G.~C.}\ \bibnamefont {Yong}}, \ and\ \bibinfo
  {author} {\bibfnamefont {M.}~\bibnamefont {Zhang}},\ }\href {\doibase
  10.1103/PhysRevLett.102.062502} {\bibfield  {journal} {\bibinfo  {journal}
  {Phys. Rev. Lett.}\ }\textbf {\bibinfo {volume} {102}},\ \bibinfo {pages}
  {062502} (\bibinfo {year} {2009})}\BibitemShut {NoStop}%
\bibitem [{\citenamefont {Xu}\ \emph {et~al.}(2013)\citenamefont {Xu},
  \citenamefont {Chen}, \citenamefont {Ko}, \citenamefont {Li},\ and\
  \citenamefont {Ma}}]{Xu:2013aza}%
  \BibitemOpen
  \bibfield  {author} {\bibinfo {author} {\bibfnamefont {J.}~\bibnamefont
  {Xu}}, \bibinfo {author} {\bibfnamefont {L.~W.}\ \bibnamefont {Chen}},
  \bibinfo {author} {\bibfnamefont {C.~M.}\ \bibnamefont {Ko}}, \bibinfo
  {author} {\bibfnamefont {B.~A.}\ \bibnamefont {Li}}, \ and\ \bibinfo {author}
  {\bibfnamefont {Y.~G.}\ \bibnamefont {Ma}},\ }\href {\doibase
  10.1103/PhysRevC.87.067601} {\bibfield  {journal} {\bibinfo  {journal} {Phys.
  Rev. C}\ }\textbf {\bibinfo {volume} {87}},\ \bibinfo {pages} {067601}
  (\bibinfo {year} {2013})}\BibitemShut {NoStop}%
\bibitem [{\citenamefont {Song}\ and\ \citenamefont {Ko}(2015)}]{Song:2015hua}%
  \BibitemOpen
  \bibfield  {author} {\bibinfo {author} {\bibfnamefont {T.}~\bibnamefont
  {Song}}\ and\ \bibinfo {author} {\bibfnamefont {C.~M.}\ \bibnamefont {Ko}},\
  }\href {\doibase 10.1103/PhysRevC.91.014901} {\bibfield  {journal} {\bibinfo
  {journal} {Phys. Rev. C}\ }\textbf {\bibinfo {volume} {91}},\ \bibinfo
  {pages} {014901} (\bibinfo {year} {2015})}\BibitemShut {NoStop}%
\bibitem [{\citenamefont {Godbey}\ \emph {et~al.}(2022)\citenamefont {Godbey},
  \citenamefont {Zhang}, \citenamefont {Holt},\ and\ \citenamefont
  {Ko}}]{Godbey:2021tbt}%
  \BibitemOpen
  \bibfield  {author} {\bibinfo {author} {\bibfnamefont {K.}~\bibnamefont
  {Godbey}}, \bibinfo {author} {\bibfnamefont {Z.}~\bibnamefont {Zhang}},
  \bibinfo {author} {\bibfnamefont {J.~W.}\ \bibnamefont {Holt}}, \ and\
  \bibinfo {author} {\bibfnamefont {C.~M.}\ \bibnamefont {Ko}},\ }\href
  {\doibase 10.1016/j.physletb.2022.137134} {\bibfield  {journal} {\bibinfo
  {journal} {Phys. Lett. B}\ }\textbf {\bibinfo {volume} {829}},\ \bibinfo
  {pages} {137134} (\bibinfo {year} {2022})}\BibitemShut {NoStop}%
\bibitem [{\citenamefont {Luong}(2024)}]{Luong:2024eaq}%
  \BibitemOpen
  \bibfield  {author} {\bibinfo {author} {\bibfnamefont {V.~B.}\ \bibnamefont
  {Luong}} (\bibinfo {collaboration} {for the STAR Collaboration}),\ }\href
  {\doibase 10.1134/S1063779624700278} {\bibfield  {journal} {\bibinfo
  {journal} {Phys. Part. Nucl.}\ }\textbf {\bibinfo {volume} {55}},\ \bibinfo
  {pages} {822} (\bibinfo {year} {2024})}\BibitemShut {NoStop}%
\bibitem [{\citenamefont {Li}\ \emph {et~al.}(2023{\natexlab{a}})\citenamefont
  {Li}, \citenamefont {Wang}, \citenamefont {Li},\ and\ \citenamefont
  {Zhang}}]{Li:2022icu}%
  \BibitemOpen
  \bibfield  {author} {\bibinfo {author} {\bibfnamefont {P.~C.}\ \bibnamefont
  {Li}}, \bibinfo {author} {\bibfnamefont {Y.~J.}\ \bibnamefont {Wang}},
  \bibinfo {author} {\bibfnamefont {Q.~F.}\ \bibnamefont {Li}}, \ and\ \bibinfo
  {author} {\bibfnamefont {H.~F.}\ \bibnamefont {Zhang}},\ }\href {\doibase
  10.1007/s11433-022-2026-5} {\bibfield  {journal} {\bibinfo  {journal} {Sci.
  China Phys. Mech. Astron.}\ }\textbf {\bibinfo {volume} {66}},\ \bibinfo
  {pages} {222011} (\bibinfo {year} {2023}{\natexlab{a}})}\BibitemShut
  {NoStop}%
\bibitem [{\citenamefont {Li}\ \emph {et~al.}(2023{\natexlab{b}})\citenamefont
  {Li}, \citenamefont {Steinheimer}, \citenamefont {Reichert}, \citenamefont
  {Kittiratpattana}, \citenamefont {Bleicher},\ and\ \citenamefont
  {Li}}]{Li:2022iil}%
  \BibitemOpen
  \bibfield  {author} {\bibinfo {author} {\bibfnamefont {P.~C.}\ \bibnamefont
  {Li}}, \bibinfo {author} {\bibfnamefont {J.}~\bibnamefont {Steinheimer}},
  \bibinfo {author} {\bibfnamefont {T.}~\bibnamefont {Reichert}}, \bibinfo
  {author} {\bibfnamefont {A.}~\bibnamefont {Kittiratpattana}}, \bibinfo
  {author} {\bibfnamefont {M.}~\bibnamefont {Bleicher}}, \ and\ \bibinfo
  {author} {\bibfnamefont {Q.~F.}\ \bibnamefont {Li}},\ }\href {\doibase
  10.1007/s11433-022-2041-8} {\bibfield  {journal} {\bibinfo  {journal} {Sci.
  China Phys. Mech. Astron.}\ }\textbf {\bibinfo {volume} {66}},\ \bibinfo
  {pages} {232011} (\bibinfo {year} {2023}{\natexlab{b}})}\BibitemShut
  {NoStop}%
\bibitem [{\citenamefont {Jhang}\ \emph {et~al.}(2021)\citenamefont {Jhang}
  \emph {et~al.}}]{SpiRIT:2020sfn}%
  \BibitemOpen
  \bibfield  {author} {\bibinfo {author} {\bibfnamefont {G.}~\bibnamefont
  {Jhang}} \emph {et~al.} (\bibinfo {collaboration} {SpiRIT and TMEP
  Collaborations}),\ }\href {\doibase 10.1016/j.physletb.2020.136016}
  {\bibfield  {journal} {\bibinfo  {journal} {Phys. Lett. B}\ }\textbf
  {\bibinfo {volume} {813}},\ \bibinfo {pages} {136016} (\bibinfo {year}
  {2021})}\BibitemShut {NoStop}%
\bibitem [{\citenamefont {Adamczewski-Musch}\ \emph {et~al.}(2020)\citenamefont
  {Adamczewski-Musch} \emph {et~al.}}]{HADES:2020ver}%
  \BibitemOpen
  \bibfield  {author} {\bibinfo {author} {\bibfnamefont {J.}~\bibnamefont
  {Adamczewski-Musch}} \emph {et~al.} (\bibinfo {collaboration} {HADES
  Collaboration}),\ }\href {\doibase 10.1140/epja/s10050-020-00237-2}
  {\bibfield  {journal} {\bibinfo  {journal} {Eur. Phys. J. A}\ }\textbf
  {\bibinfo {volume} {56}},\ \bibinfo {pages} {259} (\bibinfo {year}
  {2020})}\BibitemShut {NoStop}%
\bibitem [{\citenamefont {Xu}\ \emph {et~al.}(2024)\citenamefont {Xu} \emph
  {et~al.}}]{TMEP:2023ifw}%
  \BibitemOpen
  \bibfield  {author} {\bibinfo {author} {\bibfnamefont {J.}~\bibnamefont {Xu}}
  \emph {et~al.} (\bibinfo {collaboration} {TMEP Collaboration}),\ }\href
  {\doibase 10.1103/PhysRevC.109.044609} {\bibfield  {journal} {\bibinfo
  {journal} {Phys. Rev. C}\ }\textbf {\bibinfo {volume} {109}},\ \bibinfo
  {pages} {044609} (\bibinfo {year} {2024})}\BibitemShut {NoStop}%
\bibitem [{\citenamefont {Li}\ and\ \citenamefont {Li}(2017)}]{Li:2016xix}%
  \BibitemOpen
  \bibfield  {author} {\bibinfo {author} {\bibfnamefont {Q.~F.}\ \bibnamefont
  {Li}}\ and\ \bibinfo {author} {\bibfnamefont {Z.~X.}\ \bibnamefont {Li}},\
  }\href {\doibase 10.1016/j.physletb.2017.09.013} {\bibfield  {journal}
  {\bibinfo  {journal} {Phys. Lett. B}\ }\textbf {\bibinfo {volume} {773}},\
  \bibinfo {pages} {557} (\bibinfo {year} {2017})}\BibitemShut {NoStop}%
\bibitem [{\citenamefont {Li}\ and\ \citenamefont {Li}(2019)}]{Li:2017pis}%
  \BibitemOpen
  \bibfield  {author} {\bibinfo {author} {\bibfnamefont {Q.~F.}\ \bibnamefont
  {Li}}\ and\ \bibinfo {author} {\bibfnamefont {Z.~X.}\ \bibnamefont {Li}},\
  }\href {\doibase 10.1007/s11433-018-9336-y} {\bibfield  {journal} {\bibinfo
  {journal} {Sci. China Phys. Mech. Astron.}\ }\textbf {\bibinfo {volume}
  {62}},\ \bibinfo {pages} {972011} (\bibinfo {year} {2019})}\BibitemShut
  {NoStop}%
\bibitem [{\citenamefont {Tong}\ \emph {et~al.}(2020)\citenamefont {Tong},
  \citenamefont {Li}, \citenamefont {Li}, \citenamefont {Wang}, \citenamefont
  {Li},\ and\ \citenamefont {Liu}}]{Tong:2020dku}%
  \BibitemOpen
  \bibfield  {author} {\bibinfo {author} {\bibfnamefont {L.~Y.}\ \bibnamefont
  {Tong}}, \bibinfo {author} {\bibfnamefont {P.~C.}\ \bibnamefont {Li}},
  \bibinfo {author} {\bibfnamefont {F.~P.}\ \bibnamefont {Li}}, \bibinfo
  {author} {\bibfnamefont {Y.~J.}\ \bibnamefont {Wang}}, \bibinfo {author}
  {\bibfnamefont {Q.~F.}\ \bibnamefont {Li}}, \ and\ \bibinfo {author}
  {\bibfnamefont {F.~X.}\ \bibnamefont {Liu}},\ }\href {\doibase
  10.1088/1674-1137/44/7/074103} {\bibfield  {journal} {\bibinfo  {journal}
  {Chin. Phys. C}\ }\textbf {\bibinfo {volume} {44}},\ \bibinfo {pages}
  {074101} (\bibinfo {year} {2020})}\BibitemShut {NoStop}%
\bibitem [{\citenamefont {Larionov}\ \emph {et~al.}(2001)\citenamefont
  {Larionov}, \citenamefont {Cassing}, \citenamefont {Leupold},\ and\
  \citenamefont {Mosel}}]{Larionov:2001va}%
  \BibitemOpen
  \bibfield  {author} {\bibinfo {author} {\bibfnamefont {A.~B.}\ \bibnamefont
  {Larionov}}, \bibinfo {author} {\bibfnamefont {W.}~\bibnamefont {Cassing}},
  \bibinfo {author} {\bibfnamefont {S.}~\bibnamefont {Leupold}}, \ and\
  \bibinfo {author} {\bibfnamefont {U.}~\bibnamefont {Mosel}},\ }\href
  {\doibase 10.1016/S0375-9474(01)01216-7} {\bibfield  {journal} {\bibinfo
  {journal} {Nucl. Phys. A}\ }\textbf {\bibinfo {volume} {696}},\ \bibinfo
  {pages} {747} (\bibinfo {year} {2001})}\BibitemShut {NoStop}%
\bibitem [{\citenamefont {Oset}\ and\ \citenamefont
  {Salcedo}(1987)}]{Oset:1987re}%
  \BibitemOpen
  \bibfield  {author} {\bibinfo {author} {\bibfnamefont {E.}~\bibnamefont
  {Oset}}\ and\ \bibinfo {author} {\bibfnamefont {L.~L.}\ \bibnamefont
  {Salcedo}},\ }\href {\doibase 10.1016/0375-9474(87)90185-0} {\bibfield
  {journal} {\bibinfo  {journal} {Nucl. Phys. A}\ }\textbf {\bibinfo {volume}
  {468}},\ \bibinfo {pages} {631} (\bibinfo {year} {1987})}\BibitemShut
  {NoStop}%
\bibitem [{\citenamefont {Bohnet}\ \emph {et~al.}(1989)\citenamefont {Bohnet},
  \citenamefont {Ohtsuka}, \citenamefont {Aichelin}, \citenamefont {Linden},\
  and\ \citenamefont {Faessler}}]{Bohnet:1989dzk}%
  \BibitemOpen
  \bibfield  {author} {\bibinfo {author} {\bibfnamefont {A.}~\bibnamefont
  {Bohnet}}, \bibinfo {author} {\bibfnamefont {N.}~\bibnamefont {Ohtsuka}},
  \bibinfo {author} {\bibfnamefont {J.}~\bibnamefont {Aichelin}}, \bibinfo
  {author} {\bibfnamefont {R.}~\bibnamefont {Linden}}, \ and\ \bibinfo {author}
  {\bibfnamefont {A.}~\bibnamefont {Faessler}},\ }\href {\doibase
  10.1016/0375-9474(89)90028-6} {\bibfield  {journal} {\bibinfo  {journal}
  {Nucl. Phys. A}\ }\textbf {\bibinfo {volume} {494}},\ \bibinfo {pages} {349}
  (\bibinfo {year} {1989})}\BibitemShut {NoStop}%
\bibitem [{\citenamefont {Han}\ \emph {et~al.}(2022)\citenamefont {Han},
  \citenamefont {Shang}, \citenamefont {Zuo}, \citenamefont {Yong},\ and\
  \citenamefont {Gao}}]{Han:2022quc}%
  \BibitemOpen
  \bibfield  {author} {\bibinfo {author} {\bibfnamefont {S.~C.}\ \bibnamefont
  {Han}}, \bibinfo {author} {\bibfnamefont {X.~L.}\ \bibnamefont {Shang}},
  \bibinfo {author} {\bibfnamefont {W.}~\bibnamefont {Zuo}}, \bibinfo {author}
  {\bibfnamefont {G.~C.}\ \bibnamefont {Yong}}, \ and\ \bibinfo {author}
  {\bibfnamefont {Y.}~\bibnamefont {Gao}},\ }\href {\doibase
  10.1103/PhysRevC.106.064332} {\bibfield  {journal} {\bibinfo  {journal}
  {Phys. Rev. C}\ }\textbf {\bibinfo {volume} {106}},\ \bibinfo {pages}
  {064332} (\bibinfo {year} {2022})}\BibitemShut {NoStop}%
\bibitem [{\citenamefont {Li}\ and\ \citenamefont
  {Machleidt}(1994)}]{Li:1993ef}%
  \BibitemOpen
  \bibfield  {author} {\bibinfo {author} {\bibfnamefont {G.~Q.}\ \bibnamefont
  {Li}}\ and\ \bibinfo {author} {\bibfnamefont {R.}~\bibnamefont {Machleidt}},\
  }\href {\doibase 10.1103/PhysRevC.49.566} {\bibfield  {journal} {\bibinfo
  {journal} {Phys. Rev. C}\ }\textbf {\bibinfo {volume} {49}},\ \bibinfo
  {pages} {566} (\bibinfo {year} {1994})}\BibitemShut {NoStop}%
\bibitem [{\citenamefont {Ter~Haar}\ and\ \citenamefont
  {Malfliet}(1987)}]{TerHaar:1987ce}%
  \BibitemOpen
  \bibfield  {author} {\bibinfo {author} {\bibfnamefont {B.}~\bibnamefont
  {Ter~Haar}}\ and\ \bibinfo {author} {\bibfnamefont {R.}~\bibnamefont
  {Malfliet}},\ }\href {\doibase 10.1103/PhysRevC.36.1611} {\bibfield
  {journal} {\bibinfo  {journal} {Phys. Rev. C}\ }\textbf {\bibinfo {volume}
  {36}},\ \bibinfo {pages} {1611} (\bibinfo {year} {1987})}\BibitemShut
  {NoStop}%
\bibitem [{\citenamefont {Pandharipande}\ and\ \citenamefont
  {Pieper}(1992)}]{Pandharipande:1992zz}%
  \BibitemOpen
  \bibfield  {author} {\bibinfo {author} {\bibfnamefont {V.~R.}\ \bibnamefont
  {Pandharipande}}\ and\ \bibinfo {author} {\bibfnamefont {S.~C.}\ \bibnamefont
  {Pieper}},\ }\href {\doibase 10.1103/PhysRevC.45.791} {\bibfield  {journal}
  {\bibinfo  {journal} {Phys. Rev. C}\ }\textbf {\bibinfo {volume} {45}},\
  \bibinfo {pages} {791} (\bibinfo {year} {1992})}\BibitemShut {NoStop}%
\bibitem [{\citenamefont {Cui}\ \emph {et~al.}(2018)\citenamefont {Cui},
  \citenamefont {Zhang},\ and\ \citenamefont {Li}}]{Cui:2018gbe}%
  \BibitemOpen
  \bibfield  {author} {\bibinfo {author} {\bibfnamefont {Y.}~\bibnamefont
  {Cui}}, \bibinfo {author} {\bibfnamefont {Y.~X.}\ \bibnamefont {Zhang}}, \
  and\ \bibinfo {author} {\bibfnamefont {Z.~X.}\ \bibnamefont {Li}},\ }\href
  {\doibase 10.1103/PhysRevC.98.054605} {\bibfield  {journal} {\bibinfo
  {journal} {Phys. Rev. C}\ }\textbf {\bibinfo {volume} {98}},\ \bibinfo
  {pages} {054605} (\bibinfo {year} {2018})}\BibitemShut {NoStop}%
\bibitem [{\citenamefont {Huber}\ and\ \citenamefont
  {Aichelin}(1994)}]{Huber:1994ee}%
  \BibitemOpen
  \bibfield  {author} {\bibinfo {author} {\bibfnamefont {S.}~\bibnamefont
  {Huber}}\ and\ \bibinfo {author} {\bibfnamefont {J.}~\bibnamefont
  {Aichelin}},\ }\href {\doibase 10.1016/0375-9474(94)90232-1} {\bibfield
  {journal} {\bibinfo  {journal} {Nucl. Phys. A}\ }\textbf {\bibinfo {volume}
  {573}},\ \bibinfo {pages} {587} (\bibinfo {year} {1994})}\BibitemShut
  {NoStop}%
\bibitem [{\citenamefont {Machleidt}\ \emph {et~al.}(1987)\citenamefont
  {Machleidt}, \citenamefont {Holinde},\ and\ \citenamefont
  {Elster}}]{Machleidt:1987hj}%
  \BibitemOpen
  \bibfield  {author} {\bibinfo {author} {\bibfnamefont {R.}~\bibnamefont
  {Machleidt}}, \bibinfo {author} {\bibfnamefont {K.}~\bibnamefont {Holinde}},
  \ and\ \bibinfo {author} {\bibfnamefont {C.}~\bibnamefont {Elster}},\ }\href
  {\doibase 10.1016/S0370-1573(87)80002-9} {\bibfield  {journal} {\bibinfo
  {journal} {Phys. Rept.}\ }\textbf {\bibinfo {volume} {149}},\ \bibinfo
  {pages} {1} (\bibinfo {year} {1987})}\BibitemShut {NoStop}%
\bibitem [{\citenamefont {Larionov}\ and\ \citenamefont
  {Mosel}(2003)}]{Larionov:2003av}%
  \BibitemOpen
  \bibfield  {author} {\bibinfo {author} {\bibfnamefont {A.~B.}\ \bibnamefont
  {Larionov}}\ and\ \bibinfo {author} {\bibfnamefont {U.}~\bibnamefont
  {Mosel}},\ }\href {\doibase 10.1016/j.nuclphysa.2003.08.005} {\bibfield
  {journal} {\bibinfo  {journal} {Nucl. Phys. A}\ }\textbf {\bibinfo {volume}
  {728}},\ \bibinfo {pages} {135} (\bibinfo {year} {2003})}\BibitemShut
  {NoStop}%
\bibitem [{\citenamefont {Li}\ \emph {et~al.}(2018{\natexlab{a}})\citenamefont
  {Li}, \citenamefont {Wang}, \citenamefont {Li}, \citenamefont {Guo},\ and\
  \citenamefont {Zhang}}]{Li:2018wpv}%
  \BibitemOpen
  \bibfield  {author} {\bibinfo {author} {\bibfnamefont {P.~C.}\ \bibnamefont
  {Li}}, \bibinfo {author} {\bibfnamefont {Y.~J.}\ \bibnamefont {Wang}},
  \bibinfo {author} {\bibfnamefont {Q.~F.}\ \bibnamefont {Li}}, \bibinfo
  {author} {\bibfnamefont {C.~C.}\ \bibnamefont {Guo}}, \ and\ \bibinfo
  {author} {\bibfnamefont {H.~F.}\ \bibnamefont {Zhang}},\ }\href {\doibase
  10.1103/PhysRevC.97.044620} {\bibfield  {journal} {\bibinfo  {journal} {Phys.
  Rev. C}\ }\textbf {\bibinfo {volume} {97}},\ \bibinfo {pages} {044620}
  (\bibinfo {year} {2018}{\natexlab{a}})}\BibitemShut {NoStop}%
\bibitem [{\citenamefont {Li}\ \emph {et~al.}(2022)\citenamefont {Li},
  \citenamefont {Wang}, \citenamefont {Li},\ and\ \citenamefont
  {Zhang}}]{Li:2022wvu}%
  \BibitemOpen
  \bibfield  {author} {\bibinfo {author} {\bibfnamefont {P.~C.}\ \bibnamefont
  {Li}}, \bibinfo {author} {\bibfnamefont {Y.~J.}\ \bibnamefont {Wang}},
  \bibinfo {author} {\bibfnamefont {Q.~F.}\ \bibnamefont {Li}}, \ and\ \bibinfo
  {author} {\bibfnamefont {H.~F.}\ \bibnamefont {Zhang}},\ }\href {\doibase
  10.1016/j.physletb.2022.137019} {\bibfield  {journal} {\bibinfo  {journal}
  {Phys. Lett. B}\ }\textbf {\bibinfo {volume} {828}},\ \bibinfo {pages}
  {137019} (\bibinfo {year} {2022})}\BibitemShut {NoStop}%
\bibitem [{\citenamefont {Wang}\ \emph {et~al.}(2020)\citenamefont {Wang},
  \citenamefont {Zhang}, \citenamefont {Chen}, \citenamefont {Ko},\ and\
  \citenamefont {Ma}}]{Wang:2020xgk}%
  \BibitemOpen
  \bibfield  {author} {\bibinfo {author} {\bibfnamefont {R.}~\bibnamefont
  {Wang}}, \bibinfo {author} {\bibfnamefont {Z.}~\bibnamefont {Zhang}},
  \bibinfo {author} {\bibfnamefont {L.~W.}\ \bibnamefont {Chen}}, \bibinfo
  {author} {\bibfnamefont {C.~M.}\ \bibnamefont {Ko}}, \ and\ \bibinfo {author}
  {\bibfnamefont {Y.~G.}\ \bibnamefont {Ma}},\ }\href {\doibase
  10.1016/j.physletb.2020.135532} {\bibfield  {journal} {\bibinfo  {journal}
  {Phys. Lett. B}\ }\textbf {\bibinfo {volume} {807}},\ \bibinfo {pages}
  {135532} (\bibinfo {year} {2020})}\BibitemShut {NoStop}%
\bibitem [{\citenamefont {Li}\ \emph {et~al.}(2000)\citenamefont {Li},
  \citenamefont {Li},\ and\ \citenamefont {Mao}}]{Li:2000sha}%
  \BibitemOpen
  \bibfield  {author} {\bibinfo {author} {\bibfnamefont {Q.~F.}\ \bibnamefont
  {Li}}, \bibinfo {author} {\bibfnamefont {Z.~X.}\ \bibnamefont {Li}}, \ and\
  \bibinfo {author} {\bibfnamefont {G.~J.}\ \bibnamefont {Mao}},\ }\href
  {\doibase 10.1103/PhysRevC.62.014606} {\bibfield  {journal} {\bibinfo
  {journal} {Phys. Rev. C}\ }\textbf {\bibinfo {volume} {62}},\ \bibinfo
  {pages} {014606} (\bibinfo {year} {2000})}\BibitemShut {NoStop}%
\bibitem [{\citenamefont {Li}\ \emph {et~al.}(2004)\citenamefont {Li},
  \citenamefont {Li},\ and\ \citenamefont {Zhao}}]{Li:2003vd}%
  \BibitemOpen
  \bibfield  {author} {\bibinfo {author} {\bibfnamefont {Q.~F.}\ \bibnamefont
  {Li}}, \bibinfo {author} {\bibfnamefont {Z.~X.}\ \bibnamefont {Li}}, \ and\
  \bibinfo {author} {\bibfnamefont {E.~G.}\ \bibnamefont {Zhao}},\ }\href
  {\doibase 10.1103/PhysRevC.69.017601} {\bibfield  {journal} {\bibinfo
  {journal} {Phys. Rev. C}\ }\textbf {\bibinfo {volume} {69}},\ \bibinfo
  {pages} {017601} (\bibinfo {year} {2004})}\BibitemShut {NoStop}%
\bibitem [{\citenamefont {Kummer}\ \emph {et~al.}(2024)\citenamefont {Kummer},
  \citenamefont {Gallmeister},\ and\ \citenamefont {von
  Smekal}}]{Kummer:2023hvl}%
  \BibitemOpen
  \bibfield  {author} {\bibinfo {author} {\bibfnamefont {C.}~\bibnamefont
  {Kummer}}, \bibinfo {author} {\bibfnamefont {K.}~\bibnamefont {Gallmeister}},
  \ and\ \bibinfo {author} {\bibfnamefont {L.}~\bibnamefont {von Smekal}},\
  }\href {\doibase 10.1103/PhysRevC.109.054901} {\bibfield  {journal} {\bibinfo
   {journal} {Phys. Rev. C}\ }\textbf {\bibinfo {volume} {109}},\ \bibinfo
  {pages} {054901} (\bibinfo {year} {2024})}\BibitemShut {NoStop}%
\bibitem [{\citenamefont {Mao}\ \emph {et~al.}(1996)\citenamefont {Mao},
  \citenamefont {Li},\ and\ \citenamefont {Zhuo}}]{Mao:1996zz}%
  \BibitemOpen
  \bibfield  {author} {\bibinfo {author} {\bibfnamefont {G.~J.}\ \bibnamefont
  {Mao}}, \bibinfo {author} {\bibfnamefont {Z.~X.}\ \bibnamefont {Li}}, \ and\
  \bibinfo {author} {\bibfnamefont {Y.~Z.}\ \bibnamefont {Zhuo}},\ }\href
  {\doibase 10.1103/PhysRevC.53.2933} {\bibfield  {journal} {\bibinfo
  {journal} {Phys. Rev. C}\ }\textbf {\bibinfo {volume} {53}},\ \bibinfo
  {pages} {2933} (\bibinfo {year} {1996})}\BibitemShut {NoStop}%
\bibitem [{\citenamefont {Nan}\ \emph {et~al.}(2024)\citenamefont {Nan},
  \citenamefont {Li}, \citenamefont {Wang}, \citenamefont {Li},\ and\
  \citenamefont {Zuo}}]{Nan:2023gwp}%
  \BibitemOpen
  \bibfield  {author} {\bibinfo {author} {\bibfnamefont {M.~Z.}\ \bibnamefont
  {Nan}}, \bibinfo {author} {\bibfnamefont {P.~C.}\ \bibnamefont {Li}},
  \bibinfo {author} {\bibfnamefont {Y.~J.}\ \bibnamefont {Wang}}, \bibinfo
  {author} {\bibfnamefont {Q.~F.}\ \bibnamefont {Li}}, \ and\ \bibinfo {author}
  {\bibfnamefont {W.}~\bibnamefont {Zuo}},\ }\href {\doibase
  10.1140/epja/s10050-024-01349-9} {\bibfield  {journal} {\bibinfo  {journal}
  {Eur. Phys. J. A}\ }\textbf {\bibinfo {volume} {60}},\ \bibinfo {pages} {131}
  (\bibinfo {year} {2024})}\BibitemShut {NoStop}%
\bibitem [{\citenamefont {Long}\ \emph {et~al.}()\citenamefont {Long},
  \citenamefont {Van~Giai},\ and\ \citenamefont {Meng}}]{Long:2006nc}%
  \BibitemOpen
  \bibfield  {author} {\bibinfo {author} {\bibfnamefont {W.~H.}\ \bibnamefont
  {Long}}, \bibinfo {author} {\bibfnamefont {N.}~\bibnamefont {Van~Giai}}, \
  and\ \bibinfo {author} {\bibfnamefont {J.}~\bibnamefont {Meng}},\ }\href@noop
  {} {\ }\Eprint {http://arxiv.org/abs/nucl-th/0608009} {arXiv:nucl-th/0608009}
  \BibitemShut {NoStop}%
\bibitem [{\citenamefont {Long}\ \emph {et~al.}(2010)\citenamefont {Long},
  \citenamefont {Ring}, \citenamefont {Meng}, \citenamefont {Van~Giai},\ and\
  \citenamefont {Bertulani}}]{Long:2010qc}%
  \BibitemOpen
  \bibfield  {author} {\bibinfo {author} {\bibfnamefont {W.~H.}\ \bibnamefont
  {Long}}, \bibinfo {author} {\bibfnamefont {P.}~\bibnamefont {Ring}}, \bibinfo
  {author} {\bibfnamefont {J.}~\bibnamefont {Meng}}, \bibinfo {author}
  {\bibfnamefont {N.}~\bibnamefont {Van~Giai}}, \ and\ \bibinfo {author}
  {\bibfnamefont {C.~A.}\ \bibnamefont {Bertulani}},\ }\href {\doibase
  10.1103/PhysRevC.81.031302} {\bibfield  {journal} {\bibinfo  {journal} {Phys.
  Rev. C}\ }\textbf {\bibinfo {volume} {81}},\ \bibinfo {pages} {031302}
  (\bibinfo {year} {2010})}\BibitemShut {NoStop}%
\bibitem [{\citenamefont {Kubis}\ and\ \citenamefont
  {Kutschera}(1997)}]{Kubis:1997ew}%
  \BibitemOpen
  \bibfield  {author} {\bibinfo {author} {\bibfnamefont {S.}~\bibnamefont
  {Kubis}}\ and\ \bibinfo {author} {\bibfnamefont {M.}~\bibnamefont
  {Kutschera}},\ }\href {\doibase 10.1016/S0370-2693(97)00306-7} {\bibfield
  {journal} {\bibinfo  {journal} {Phys. Lett. B}\ }\textbf {\bibinfo {volume}
  {399}},\ \bibinfo {pages} {191} (\bibinfo {year} {1997})}\BibitemShut
  {NoStop}%
\bibitem [{\citenamefont {Dutra}\ \emph {et~al.}(2014)\citenamefont {Dutra},
  \citenamefont {Louren\c{c}o}, \citenamefont {Avancini} \emph
  {et~al.}}]{Dutra:2014qga}%
  \BibitemOpen
  \bibfield  {author} {\bibinfo {author} {\bibfnamefont {M.}~\bibnamefont
  {Dutra}}, \bibinfo {author} {\bibfnamefont {O.}~\bibnamefont {Louren\c{c}o}},
  \bibinfo {author} {\bibfnamefont {S.~S.}\ \bibnamefont {Avancini}},  \emph
  {et~al.},\ }\href {\doibase 10.1103/PhysRevC.90.055203} {\bibfield  {journal}
  {\bibinfo  {journal} {Phys. Rev. C}\ }\textbf {\bibinfo {volume} {90}},\
  \bibinfo {pages} {055203} (\bibinfo {year} {2014})}\BibitemShut {NoStop}%
\bibitem [{\citenamefont {Santos}\ \emph {et~al.}()\citenamefont {Santos},
  \citenamefont {Malik},\ and\ \citenamefont {Provid\^encia}}]{Santos:2024aii}%
  \BibitemOpen
  \bibfield  {author} {\bibinfo {author} {\bibfnamefont {L.~G. T.~d.}\
  \bibnamefont {Santos}}, \bibinfo {author} {\bibfnamefont {T.}~\bibnamefont
  {Malik}}, \ and\ \bibinfo {author} {\bibfnamefont {C.}~\bibnamefont
  {Provid\^encia}},\ }\href@noop {} {\ }\Eprint
  {http://arxiv.org/abs/2412.04946} {arXiv:2412.04946 [nucl-th]} \BibitemShut
  {NoStop}%
\bibitem [{\citenamefont {Roca-Maza}\ \emph {et~al.}(2011)\citenamefont
  {Roca-Maza}, \citenamefont {Vinas}, \citenamefont {Centelles}, \citenamefont
  {Ring},\ and\ \citenamefont {Schuck}}]{Roca-Maza:2011alv}%
  \BibitemOpen
  \bibfield  {author} {\bibinfo {author} {\bibfnamefont {X.}~\bibnamefont
  {Roca-Maza}}, \bibinfo {author} {\bibfnamefont {X.}~\bibnamefont {Vinas}},
  \bibinfo {author} {\bibfnamefont {M.}~\bibnamefont {Centelles}}, \bibinfo
  {author} {\bibfnamefont {P.}~\bibnamefont {Ring}}, \ and\ \bibinfo {author}
  {\bibfnamefont {P.}~\bibnamefont {Schuck}},\ }\href {\doibase
  10.1103/PhysRevC.84.054309} {\bibfield  {journal} {\bibinfo  {journal} {Phys.
  Rev. C}\ }\textbf {\bibinfo {volume} {84}},\ \bibinfo {pages} {054309}
  (\bibinfo {year} {2011})},\ \bibinfo {note} {[Erratum: Phys.Rev.C 93, 069905
  (2016)]}\BibitemShut {NoStop}%
\bibitem [{\citenamefont {Miyatsu}\ \emph {et~al.}(2022)\citenamefont
  {Miyatsu}, \citenamefont {Cheoun},\ and\ \citenamefont
  {Saito}}]{Miyatsu:2022wuy}%
  \BibitemOpen
  \bibfield  {author} {\bibinfo {author} {\bibfnamefont {T.}~\bibnamefont
  {Miyatsu}}, \bibinfo {author} {\bibfnamefont {M.~K.}\ \bibnamefont {Cheoun}},
  \ and\ \bibinfo {author} {\bibfnamefont {K.}~\bibnamefont {Saito}},\ }\href
  {\doibase 10.3847/1538-4357/ac5f40} {\bibfield  {journal} {\bibinfo
  {journal} {Astrophys. J.}\ }\textbf {\bibinfo {volume} {929}},\ \bibinfo
  {pages} {82} (\bibinfo {year} {2022})}\BibitemShut {NoStop}%
\bibitem [{\citenamefont {Buss}\ \emph {et~al.}(2012)\citenamefont {Buss},
  \citenamefont {Gaitanos}, \citenamefont {Gallmeister}, \citenamefont {van
  Hees}, \citenamefont {Kaskulov}, \citenamefont {Lalakulich}, \citenamefont
  {Larionov}, \citenamefont {Leitner}, \citenamefont {Weil},\ and\
  \citenamefont {Mosel}}]{Buss:2011mx}%
  \BibitemOpen
  \bibfield  {author} {\bibinfo {author} {\bibfnamefont {O.}~\bibnamefont
  {Buss}}, \bibinfo {author} {\bibfnamefont {T.}~\bibnamefont {Gaitanos}},
  \bibinfo {author} {\bibfnamefont {K.}~\bibnamefont {Gallmeister}}, \bibinfo
  {author} {\bibfnamefont {H.}~\bibnamefont {van Hees}}, \bibinfo {author}
  {\bibfnamefont {M.}~\bibnamefont {Kaskulov}}, \bibinfo {author}
  {\bibfnamefont {O.}~\bibnamefont {Lalakulich}}, \bibinfo {author}
  {\bibfnamefont {A.~B.}\ \bibnamefont {Larionov}}, \bibinfo {author}
  {\bibfnamefont {T.}~\bibnamefont {Leitner}}, \bibinfo {author} {\bibfnamefont
  {J.}~\bibnamefont {Weil}}, \ and\ \bibinfo {author} {\bibfnamefont
  {U.}~\bibnamefont {Mosel}},\ }\href {\doibase 10.1016/j.physrep.2011.12.001}
  {\bibfield  {journal} {\bibinfo  {journal} {Phys. Rept.}\ }\textbf {\bibinfo
  {volume} {512}},\ \bibinfo {pages} {1} (\bibinfo {year} {2012})}\BibitemShut
  {NoStop}%
\bibitem [{\citenamefont {Hofmann}\ \emph {et~al.}(2001)\citenamefont
  {Hofmann}, \citenamefont {Keil},\ and\ \citenamefont
  {Lenske}}]{Hofmann:2000vz}%
  \BibitemOpen
  \bibfield  {author} {\bibinfo {author} {\bibfnamefont {F.}~\bibnamefont
  {Hofmann}}, \bibinfo {author} {\bibfnamefont {C.~M.}\ \bibnamefont {Keil}}, \
  and\ \bibinfo {author} {\bibfnamefont {H.}~\bibnamefont {Lenske}},\ }\href
  {\doibase 10.1103/PhysRevC.64.034314} {\bibfield  {journal} {\bibinfo
  {journal} {Phys. Rev. C}\ }\textbf {\bibinfo {volume} {64}},\ \bibinfo
  {pages} {034314} (\bibinfo {year} {2001})}\BibitemShut {NoStop}%
\bibitem [{\citenamefont {Li}\ \emph {et~al.}(2018{\natexlab{b}})\citenamefont
  {Li}, \citenamefont {Sedrakian},\ and\ \citenamefont {Weber}}]{Li:2018qaw}%
  \BibitemOpen
  \bibfield  {author} {\bibinfo {author} {\bibfnamefont {J.~J.}\ \bibnamefont
  {Li}}, \bibinfo {author} {\bibfnamefont {A.}~\bibnamefont {Sedrakian}}, \
  and\ \bibinfo {author} {\bibfnamefont {F.}~\bibnamefont {Weber}},\ }\href
  {\doibase 10.1016/j.physletb.2018.06.051} {\bibfield  {journal} {\bibinfo
  {journal} {Phys. Lett. B}\ }\textbf {\bibinfo {volume} {783}},\ \bibinfo
  {pages} {234} (\bibinfo {year} {2018}{\natexlab{b}})}\BibitemShut {NoStop}%
\bibitem [{\citenamefont {Kosov}\ \emph {et~al.}(1998)\citenamefont {Kosov},
  \citenamefont {Fuchs}, \citenamefont {Martemyanov},\ and\ \citenamefont
  {Faessler}}]{Kosov:1998gp}%
  \BibitemOpen
  \bibfield  {author} {\bibinfo {author} {\bibfnamefont {D.~S.}\ \bibnamefont
  {Kosov}}, \bibinfo {author} {\bibfnamefont {C.}~\bibnamefont {Fuchs}},
  \bibinfo {author} {\bibfnamefont {B.~V.}\ \bibnamefont {Martemyanov}}, \ and\
  \bibinfo {author} {\bibfnamefont {A.}~\bibnamefont {Faessler}},\ }\href
  {\doibase 10.1016/S0370-2693(97)01598-0} {\bibfield  {journal} {\bibinfo
  {journal} {Phys. Lett. B}\ }\textbf {\bibinfo {volume} {421}},\ \bibinfo
  {pages} {37} (\bibinfo {year} {1998})}\BibitemShut {NoStop}%
\bibitem [{\citenamefont {Sun}\ \emph {et~al.}(2019)\citenamefont {Sun},
  \citenamefont {Zhang}, \citenamefont {Zhang},\ and\ \citenamefont
  {Xia}}]{Sun:2018tmw}%
  \BibitemOpen
  \bibfield  {author} {\bibinfo {author} {\bibfnamefont {T.~T.}\ \bibnamefont
  {Sun}}, \bibinfo {author} {\bibfnamefont {S.~S.}\ \bibnamefont {Zhang}},
  \bibinfo {author} {\bibfnamefont {Q.~L.}\ \bibnamefont {Zhang}}, \ and\
  \bibinfo {author} {\bibfnamefont {C.~J.}\ \bibnamefont {Xia}},\ }\href
  {\doibase 10.1103/PhysRevD.99.023004} {\bibfield  {journal} {\bibinfo
  {journal} {Phys. Rev. D}\ }\textbf {\bibinfo {volume} {99}},\ \bibinfo
  {pages} {023004} (\bibinfo {year} {2019})}\BibitemShut {NoStop}%
\bibitem [{\citenamefont {Drago}\ \emph {et~al.}(2014)\citenamefont {Drago},
  \citenamefont {Lavagno}, \citenamefont {Pagliara},\ and\ \citenamefont
  {Pigato}}]{Drago:2014oja}%
  \BibitemOpen
  \bibfield  {author} {\bibinfo {author} {\bibfnamefont {A.}~\bibnamefont
  {Drago}}, \bibinfo {author} {\bibfnamefont {A.}~\bibnamefont {Lavagno}},
  \bibinfo {author} {\bibfnamefont {G.}~\bibnamefont {Pagliara}}, \ and\
  \bibinfo {author} {\bibfnamefont {D.}~\bibnamefont {Pigato}},\ }\href
  {\doibase 10.1103/PhysRevC.90.065809} {\bibfield  {journal} {\bibinfo
  {journal} {Phys. Rev. C}\ }\textbf {\bibinfo {volume} {90}},\ \bibinfo
  {pages} {065809} (\bibinfo {year} {2014})}\BibitemShut {NoStop}%
\bibitem [{\citenamefont {Wehrberger}\ \emph {et~al.}(1989)\citenamefont
  {Wehrberger}, \citenamefont {Bedau},\ and\ \citenamefont
  {Beck}}]{Wehrberger:1989cd}%
  \BibitemOpen
  \bibfield  {author} {\bibinfo {author} {\bibfnamefont {K.}~\bibnamefont
  {Wehrberger}}, \bibinfo {author} {\bibfnamefont {C.}~\bibnamefont {Bedau}}, \
  and\ \bibinfo {author} {\bibfnamefont {F.}~\bibnamefont {Beck}},\ }\href
  {\doibase 10.1016/0375-9474(89)90008-0} {\bibfield  {journal} {\bibinfo
  {journal} {Nucl. Phys. A}\ }\textbf {\bibinfo {volume} {504}},\ \bibinfo
  {pages} {797} (\bibinfo {year} {1989})}\BibitemShut {NoStop}%
\bibitem [{\citenamefont {Raduta}(2021)}]{Raduta:2021xiz}%
  \BibitemOpen
  \bibfield  {author} {\bibinfo {author} {\bibfnamefont {A.~R.}\ \bibnamefont
  {Raduta}},\ }\href {\doibase 10.1016/j.physletb.2021.136070} {\bibfield
  {journal} {\bibinfo  {journal} {Phys. Lett. B}\ }\textbf {\bibinfo {volume}
  {814}},\ \bibinfo {pages} {136070} (\bibinfo {year} {2021})}\BibitemShut
  {NoStop}%
\bibitem [{\citenamefont {Li}\ \emph {et~al.}(1991)\citenamefont {Li},
  \citenamefont {Bauer},\ and\ \citenamefont {Bertsch}}]{Li:1991pq}%
  \BibitemOpen
  \bibfield  {author} {\bibinfo {author} {\bibfnamefont {B.~A.}\ \bibnamefont
  {Li}}, \bibinfo {author} {\bibfnamefont {W.}~\bibnamefont {Bauer}}, \ and\
  \bibinfo {author} {\bibfnamefont {G.~F.}\ \bibnamefont {Bertsch}},\ }\href
  {\doibase 10.1103/PhysRevC.44.2095} {\bibfield  {journal} {\bibinfo
  {journal} {Phys. Rev. C}\ }\textbf {\bibinfo {volume} {44}},\ \bibinfo
  {pages} {2095} (\bibinfo {year} {1991})}\BibitemShut {NoStop}%
\bibitem [{\citenamefont {Wang}\ \emph {et~al.}(1991)\citenamefont {Wang},
  \citenamefont {Li}, \citenamefont {Bauer},\ and\ \citenamefont
  {Randrup}}]{Wang:1991sj}%
  \BibitemOpen
  \bibfield  {author} {\bibinfo {author} {\bibfnamefont {S.~J.}\ \bibnamefont
  {Wang}}, \bibinfo {author} {\bibfnamefont {B.~A.}\ \bibnamefont {Li}},
  \bibinfo {author} {\bibfnamefont {W.}~\bibnamefont {Bauer}}, \ and\ \bibinfo
  {author} {\bibfnamefont {J.}~\bibnamefont {Randrup}},\ }\href {\doibase
  10.1016/0003-4916(91)90031-3} {\bibfield  {journal} {\bibinfo  {journal}
  {Annals Phys.}\ }\textbf {\bibinfo {volume} {209}},\ \bibinfo {pages} {251}
  (\bibinfo {year} {1991})}\BibitemShut {NoStop}%
\bibitem [{\citenamefont {Almaalol}\ \emph {et~al.}()\citenamefont {Almaalol},
  \citenamefont {Hippert}, \citenamefont {Noronha-Hostler} \emph
  {et~al.}}]{Almaalol:2022xwv}%
  \BibitemOpen
  \bibfield  {author} {\bibinfo {author} {\bibfnamefont {D.}~\bibnamefont
  {Almaalol}}, \bibinfo {author} {\bibfnamefont {M.}~\bibnamefont {Hippert}},
  \bibinfo {author} {\bibfnamefont {J.}~\bibnamefont {Noronha-Hostler}},  \emph
  {et~al.},\ }\href@noop {} {\ }\Eprint
  {http://arxiv.org/abs/nucl-ex/2209.05009} {arXiv:nucl-ex/2209.05009}
  \BibitemShut {NoStop}%
\bibitem [{\citenamefont {Li}\ \emph {et~al.}(2023{\natexlab{c}})\citenamefont
  {Li}, \citenamefont {Yong},\ and\ \citenamefont {Zhang}}]{Li:2022cfd}%
  \BibitemOpen
  \bibfield  {author} {\bibinfo {author} {\bibfnamefont {A.}~\bibnamefont
  {Li}}, \bibinfo {author} {\bibfnamefont {G.~C.}\ \bibnamefont {Yong}}, \ and\
  \bibinfo {author} {\bibfnamefont {Y.~X.}\ \bibnamefont {Zhang}},\ }\href
  {\doibase 10.1103/PhysRevD.107.043005} {\bibfield  {journal} {\bibinfo
  {journal} {Phys. Rev. D}\ }\textbf {\bibinfo {volume} {107}},\ \bibinfo
  {pages} {043005} (\bibinfo {year} {2023}{\natexlab{c}})}\BibitemShut
  {NoStop}%
\bibitem [{\citenamefont {Zhang}\ and\ \citenamefont
  {Ko}(2017)}]{Zhang:2017mps}%
  \BibitemOpen
  \bibfield  {author} {\bibinfo {author} {\bibfnamefont {Z.}~\bibnamefont
  {Zhang}}\ and\ \bibinfo {author} {\bibfnamefont {C.~M.}\ \bibnamefont {Ko}},\
  }\href {\doibase 10.1103/PhysRevC.95.064604} {\bibfield  {journal} {\bibinfo
  {journal} {Phys. Rev. C}\ }\textbf {\bibinfo {volume} {95}},\ \bibinfo
  {pages} {064604} (\bibinfo {year} {2017})}\BibitemShut {NoStop}%
\bibitem [{\citenamefont {Mao}\ \emph {et~al.}(1994)\citenamefont {Mao},
  \citenamefont {Li}, \citenamefont {Zhuo}, \citenamefont {Han},\ and\
  \citenamefont {Yu}}]{Mao:1994zza}%
  \BibitemOpen
  \bibfield  {author} {\bibinfo {author} {\bibfnamefont {G.~J.}\ \bibnamefont
  {Mao}}, \bibinfo {author} {\bibfnamefont {Z.~X.}\ \bibnamefont {Li}},
  \bibinfo {author} {\bibfnamefont {Y.~Z.}\ \bibnamefont {Zhuo}}, \bibinfo
  {author} {\bibfnamefont {Y.~L.}\ \bibnamefont {Han}}, \ and\ \bibinfo
  {author} {\bibfnamefont {Z.~Q.}\ \bibnamefont {Yu}},\ }\href {\doibase
  10.1103/PhysRevC.49.3137} {\bibfield  {journal} {\bibinfo  {journal} {Phys.
  Rev. C}\ }\textbf {\bibinfo {volume} {49}},\ \bibinfo {pages} {3137}
  (\bibinfo {year} {1994})}\BibitemShut {NoStop}%
\bibitem [{\citenamefont {Li}\ and\ \citenamefont {Ko}(1995)}]{Li:1995pra}%
  \BibitemOpen
  \bibfield  {author} {\bibinfo {author} {\bibfnamefont {B.~A.}\ \bibnamefont
  {Li}}\ and\ \bibinfo {author} {\bibfnamefont {C.~M.}\ \bibnamefont {Ko}},\
  }\href {\doibase 10.1103/PhysRevC.52.2037} {\bibfield  {journal} {\bibinfo
  {journal} {Phys. Rev. C}\ }\textbf {\bibinfo {volume} {52}},\ \bibinfo
  {pages} {2037} (\bibinfo {year} {1995})}\BibitemShut {NoStop}%
\bibitem [{\citenamefont {Cai}\ \emph {et~al.}(1998)\citenamefont {Cai},
  \citenamefont {Feng}, \citenamefont {Shen}, \citenamefont {Ma}, \citenamefont
  {Wang},\ and\ \citenamefont {Ye}}]{Cai:1998iv}%
  \BibitemOpen
  \bibfield  {author} {\bibinfo {author} {\bibfnamefont {X.~Z.}\ \bibnamefont
  {Cai}}, \bibinfo {author} {\bibfnamefont {J.}~\bibnamefont {Feng}}, \bibinfo
  {author} {\bibfnamefont {W.~Q.}\ \bibnamefont {Shen}}, \bibinfo {author}
  {\bibfnamefont {Y.~G.}\ \bibnamefont {Ma}}, \bibinfo {author} {\bibfnamefont
  {J.~S.}\ \bibnamefont {Wang}}, \ and\ \bibinfo {author} {\bibfnamefont
  {W.}~\bibnamefont {Ye}},\ }\href {\doibase 10.1103/PhysRevC.58.572}
  {\bibfield  {journal} {\bibinfo  {journal} {Phys. Rev. C}\ }\textbf {\bibinfo
  {volume} {58}},\ \bibinfo {pages} {572} (\bibinfo {year} {1998})}\BibitemShut
  {NoStop}%
\bibitem [{\citenamefont {Su}\ \emph {et~al.}(2016)\citenamefont {Su},
  \citenamefont {Huang}, \citenamefont {Xie},\ and\ \citenamefont
  {Zhang}}]{Su:2016adl}%
  \BibitemOpen
  \bibfield  {author} {\bibinfo {author} {\bibfnamefont {J.}~\bibnamefont
  {Su}}, \bibinfo {author} {\bibfnamefont {C.~Y.}\ \bibnamefont {Huang}},
  \bibinfo {author} {\bibfnamefont {W.~J.}\ \bibnamefont {Xie}}, \ and\
  \bibinfo {author} {\bibfnamefont {F.~S.}\ \bibnamefont {Zhang}},\ }\href
  {\doibase 10.1140/epja/i2016-16207-x} {\bibfield  {journal} {\bibinfo
  {journal} {Eur. Phys. J. A}\ }\textbf {\bibinfo {volume} {52}},\ \bibinfo
  {pages} {207} (\bibinfo {year} {2016})}\BibitemShut {NoStop}%
\end{thebibliography}%
			
\end{document}